\documentclass[fleqn,usenatbib]{mnras}
\usepackage[T1]{fontenc}
\DeclareRobustCommand{\VAN}[3]{#2}
\let\VANthebibliography\thebibliography
\def\thebibliography{\DeclareRobustCommand{\VAN}[3]{##3}\VANthebibliography}    
\usepackage[table,xcdraw]{xcolor}
\usepackage{graphicx}	
\usepackage{amsmath}	
\usepackage{amssymb}	
\usepackage{gensymb}
\usepackage{longtable}
\usepackage[utf8]{inputenc} 
\usepackage{pdflscape}
\usepackage{afterpage}
\usepackage{adjustbox}
\usepackage{placeins}
\usepackage{capt-of}
\usepackage{textcase}
\usepackage{caption}
\usepackage{array}
\usepackage{natbib}
\usepackage{multirow}
\usepackage{tabularx}
\usepackage{booktabs}
\usepackage{tablefootnote}
\usepackage{threeparttable} 
\usepackage{amsfonts}
\usepackage{newtxtext,newtxmath}
\usepackage{geometry}
\usepackage{changepage}

\DeclareUnicodeCharacter{2212}{-} 

\title[Herbig Ae/Be stars identified from LAMOST DR5]{Spectroscopic study of Herbig Ae/Be stars in the Galactic Anti-center region from LAMOST DR5}

\author[Nidhi et al.]{S. Nidhi$^{1}$\thanks{E-mail:nidhi.sabu@res.christuniversity.in}, Blesson Mathew$^{1}$, B. Shridharan$^{1}$, R. Arun$^{2}$, R. Anusha$^{3}$,
and Sreeja S. Kartha$^{1}$
\\
$^{1}$Department of Physics and Electronics, CHRIST (Deemed to be University), Bangalore 560029, India\\
$^{2}$Indian Institute of Astrophysics (IIA), Sarjapur Road, Koramangala, Bangalore 560034, India \\
$^{3}$Department of Physics \& Astronomy, University of Western Ontario, London, Ontario, N6A 5B7, Canada\\
}

\date{Accepted 2023 June 30. Received 2023 June 19; in original form 2022 February 06}

\begin{document}

\pubyear{2023}
\defcitealias{2021RAA....21..288S}{Paper 1}
\label{firstpage}
\pagerange{\pageref{firstpage}--\pageref{lastpage}}
\maketitle

\begin{abstract}
We study a sample of 119 Herbig Ae/Be stars in the Galactic anti-center direction using the spectroscopic data from Large sky Area Multi-Object fiber Spectroscopic Telescope (LAMOST) survey program. Emission lines of hydrogen belonging to the Balmer and Paschen series, and metallic lines of species such as Fe{\sc ii}, O{\sc i}, Ca{\sc ii} triplet are identified. A moderate correlation is observed between the emission strengths of H$\alpha$ and Fe{\sc ii} 5169 \AA, suggesting a possible common emission region for Fe{\sc ii} lines and one of the components of H$\alpha$. We explored a technique for the extinction correction of the HAeBe stars using diffuse interstellar bands present in the spectrum. We estimated the stellar parameters such as age and mass of these HAeBe stars, which are found to be in the range 0.1 -- 10 Myr and 1.5 -- 10 $M_{\odot}$, respectively. We found that the mass accretion rate of the HAeBe stars in the Galactic anti-center direction follows the relation $\dot{M}_{acc}$ $\propto$ $M_{*}^{3.12^{+0.21}_{-0.34}}$, which is similar to the relation derived for HAeBe stars in other regions of the Galaxy. The mass accretion rate of HAeBe stars is found to have a functional form of $\dot{M}_{acc} \propto t^{-1.1 \pm 0.2}$ with age, in agreement with previous studies. 
\end{abstract}

\begin{keywords}
stars: emission-line, HAeBe-techniques: photometric-techniques: spectroscopic-accretion
\end{keywords}


\section{Introduction}

Herbig Ae/Be stars (HAeBe) are pre-main sequence emission-line stars (ELS) having a circumstellar accretion disc \citep{1998ARA&A..36..233W}. \cite{1960ApJS....4..337H} defined this as a class of 26 Ae/Be stars in pre-main sequence (PMS) phase, which shows H$\alpha$ emission feature in their spectrum and associated with nebulosity. Later, the sample set of HAeBe stars was expanded by \cite{1972ApJ...173..353S}, \cite{1984A&AS...55..109F}, and \cite{1994A&AS..104..315T}. Further, F-type stars (F0 - F5) with Balmer emission lines were also included into the HAeBe category \citep{1998A&A...331..211M, 1992ApJS...82..285H}. Over the years, the studies on HAeBe stars have surged due to the availability of large-scale photometric and spectroscopic surveys such as Gaia \citep{2020yCat.1350....0G}, 2MASS \citep{2003yCat.2246....0C}, APOGEE \citep{2017AJ....154...94M}, and LAMOST \citep{2012RAA....12.1197C}. The emission lines, whether they belong to hydrogen or other metal species, are primarily formed in the disc/envelope around HAeBe stars. A few prominent studies were performed to understand the mechanisms for the formation and evolution of emission lines/features in HAeBe stars (e.g. H$\alpha$ \citep{1977ApJ...218..438G, 1984A&AS...55..109F}, Br$\gamma$ \citep{2016A&A...590A..97T}, [O{\sc i}] \citep{1994ApJS...93..485H, 2008A&A...485..487V}, Ca{\sc ii} triplet \citep{1992ApJS...82..285H, 1995A&A...301..155B}, He{\sc i} \citep{2011AN....332..238O}, and O{\sc i} \citep{2018ApJ...857...30M}). Emission lines belonging to Fe{\sc ii} transitions are also seen in the spectra of HAeBe stars \citep{2004AJ....127.1682H}, though it has not been studied in an explicit manner. 

The presence of dust in the circumstellar disc is confirmed from the Infrared (IR) and millimeter excess seen in the Spectral Energy Distribution (SED) of HAeBe stars \citep{1992ApJ...397..613H, 2001A&A...371..186N}. The gaseous and dusty disc around the young stellar objects (YSOs) is the resource that provides the material which continuously accretes onto the growing star \citep{2016ARA&A..54..135H}. Guided by the previous works on the estimation of accretion rates for PMS stars, a magnetically driven accretion mechanism seems to be a valid scenario for low-mass Herbig stars (HAe) and T Tauri stars (TTS), while a boundary layer accretion mechanism seems to favor high-mass Herbig Be (HBe) stars \citep{2005MNRAS.359.1049V, 2018MNRAS.474...88H, 2016A&A...592A..50S, 2017MNRAS.472..854A, 2020Galax...8...39M, 2020MNRAS.493..234W}.   

We performed a spectroscopic study of HAeBe stars identified from the fifth data release of the Large Sky Area Multi-Object Fiber Spectroscopic Telescope (LAMOST DR5). Most of the HAeBe stars identified from the literature are observed towards the direction of the Galactic center \citep{1994A&AS..104..315T, 2004AJ....127.1682H, 2021A&A...650A.182G, 2021A&A...652A.133V, 2022ApJ...930...39V}. Thanks to the observational strategy of the LAMOST survey, we obtain a new set of HAeBe stars spread towards the Galactic anti-center direction. Pioneering work on the search of PMS stars towards this region was carried out by \cite{2016RAA....16..138H}. They presented a catalog of early-type ELS from LAMOST DR2, where they identified a sample of 26 HAeBe stars. Recently, \citet{2021RAA....21..288S} (hereafter Paper 1) discovered 2716 hot ELS from LAMOST DR5, among which 56 were HAeBe stars. \cite{2022ApJS..259...38Z} presented a catalog of ELS in which 62 stars are identified as new HAeBe stars. Apart from these, there exists no considerable information about the stellar parameters and spectroscopy of HAeBe stars towards the Galactic anti-center region in the literature. Hence, by using the optical spectra from the LAMOST survey program, we are homogeneously analyzing the largest sample of HAeBe stars towards this region. This research will help to provide important information on the properties of HAeBe stars in various regions of the Galaxy, which in future can be extended beyond the Galactic regime.

The main aim of this work is to study the spectra of HAeBe stars identified from LAMOST DR5 and to determine the stellar parameters and accretion rates using a homogeneous approach. We identified a sample of 119 possible HAeBe stars from LAMOST DR5. We combined the optical and infrared photometric information from Gaia EDR3 \citep{2020yCat.1350....0G}, Two Micron All Sky Survey (2MASS; \citealp{2003yCat.2246....0C}) and Wide-field Infrared Survey Explorer (WISE; \citealp{2014yCat.2328....0C}) with the LAMOST spectroscopic data, to identify the HAeBe stars. Spectroscopic analysis is performed for the HAeBe stars, providing insights about the region of formation and of emission lines (belonging to various species) in the accretion disc of HAeBe stars. We have also explored a method for the extinction correction of the HAeBe stars. This paper is organized as follows. In Section \ref{sec:2}, we discuss the LAMOST survey program and the sample selection method employed to identify the HAeBe stars from the LAMOST catalog. Section \ref{sec:3} describes the spectral features and evolutionary status of HAeBe stars from this study. We also discuss the dependency of mass accretion rate with stellar parameters such as age and mass. The results are summarized in Section \ref{sec:4}. 
    
\section{Data procurement and identification of the sample}
\label{sec:2}

In this section, we discuss in brief about the LAMOST telescope and observation strategy, from which the data is used in this work. Also, a detailed description of the step-by-step criteria used to select a consolidated sample of HAeBe stars is given. 

\subsection{LAMOST}
\label{subsec:2.1}

The Large Sky Area Multi-Object Fiber Spectroscopic Telescope, also known as Guo Shoujing Telescope, is a quasi-meridian reflecting Schmidt telescope operated by the Chinese Academy of Sciences \citep{2012RAA....12.1197C}. The 4000 optical fibers in the focal plane allow it to obtain multiple spectra at one go, covering the wavelength range from 3800 \AA~to 9000 \AA. LAMOST has a wide Field-of-View (FoV) of 5{$^{\circ}$} and an effective aperture of 3.6m - 4.9m \citep{2012arXiv1206.3569Z}. LAMOST has observed towards the Galactic anti-center direction and obtained spectra of around 10 million objects in the sky \citep{2015MNRAS.448..855Y}. The program was initiated in 2012 and completed its five-year of phase-I survey in low-resolution mode (R$\sim$1800). LAMOST conducted galactic and extragalactic surveys, namely, LAMOST Experiment for Galactic Understanding and Exploration (LEGUE; \citealp{2012RAA....12..735D}) and LAMOST Extra GAlactic Survey (LEGAS; \citealp{2012arXiv1206.3569Z}). Later, the survey was extended to phase-II (from 2018) by combining the spectra taken in low- and medium-resolution (R$\sim$7500) modes. 

By 2017, the LAMOST DR5 \citep{2019yCat.5164....0L} catalog had accumulated 9,026,365 spectra, of which 8,183,160 were labeled as stars. The reduction of raw data was carried out by LAMOST two-dimensional (2D) pipeline, which included basic reduction steps such as de-biasing, flat-fielding, fiber-tracing, sky subtraction, and wavelength calibration \citep{2015RAA....15.1095L}. Furthermore, the LAMOST one-dimensional (1D) pipeline extracted and classified spectra into categories such as \textit{stars}, \textit{galaxies}, \textit{QSOs} and \textit{unknowns} \citep{2012arXiv1206.3569Z,2012RAA....12.1197C}. We retrieved the spectra of stars from the fifth data release of LAMOST, which is used for the present study.

\subsection{Sample Selection}
\label{subsec:2.2}

\begin{figure}
\centering
\includegraphics[width=\columnwidth]{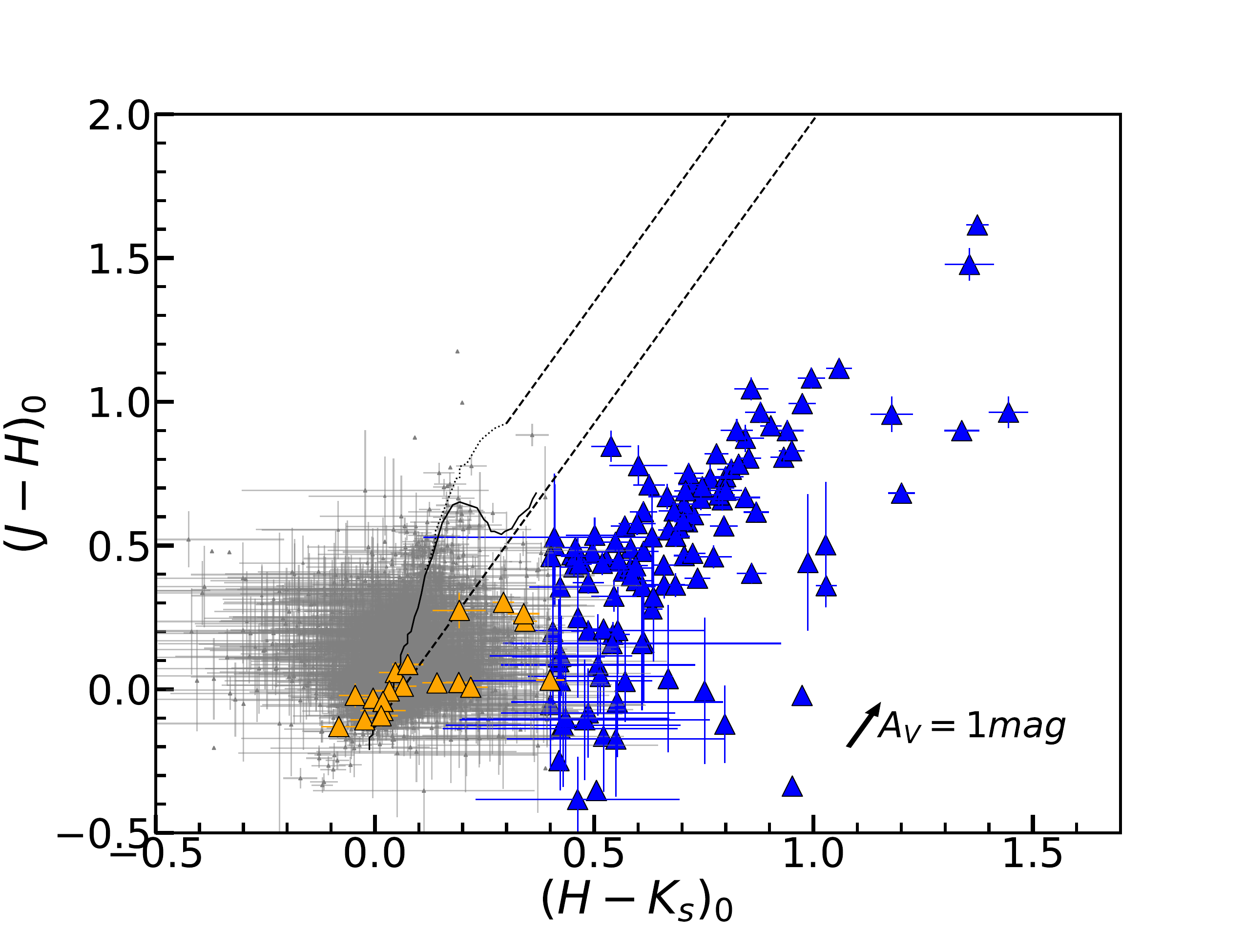}
    \caption{2MASS (\textit{H}-\textit{K}$_s$)$_0$ vs (\textit{J}-\textit{H})$_0$ color-color diagram of 3,850 ELS, shown in grey in the figure. The blue triangles represent 127 stars identified as HAeBe candidates, which satisfies the (\textit{H}-\textit{K}$_s$)$_0$ > 0.4 mag color criterion given by \protect\citet{1984A&AS...55..109F}. The orange triangles represent 24 stars identified using \textit{$n_{(K_{s}-W2)}$} > -1.5 criterion. The main-sequence, the giant branch and reddening vectors are represented by solid, dotted and dashed lines, respectively, which is adopted from \protect\cite{1983A&A...128...84K} and converted to 2MASS system using the relation from \protect\citet{2001AJ....121.2851C}. The arrow represents the direction in which a star will move for an extinction of A$_V$ = 1 mag.}
    \label{fig:CCD}
\end{figure}

The sample selection process of HAeBe stars identified from the LAMOST DR5 is discussed in \citetalias{2021RAA....21..288S}. A brief summary of the candidate selection criteria is given here, and further, we discuss the details on the completeness of the sample used in this study. The presence of H$\alpha$ emission line in the optical spectrum, which confirms the presence of the circumstellar disc or envelope, is used as the primary criterion for identifying an ELS. In \citetalias{2021RAA....21..288S}, the spectra of stars classified as B, and A types by the LAMOST stellar parameter pipeline (LASP; \citealp{2011RAA....11..924W}) were retrieved from the LAMOST DR5 archive. Using an automated python routine, spectra showing H$\alpha$ emission, peaking between 6561 \AA~and 6568 \AA~were fetched. Multi-epoch spectra with the Signal-to-Noise Ratio (SNR) in SDSS r-band less than 10 (\textit{SNR$_r$} < 10; \citealp{2016RAA....16..138H}) were removed, leaving 3339 unique ELS. Due to the inaccuracy in the spectral type provided by LAMOST DR5, the spectral type was re-estimated through template matching technique using MILES stellar library (\citealp{2006MNRAS.371..703S}; see section \ref{subsec:3.2}). By definition, the spectral type of HAeBe stars ranges from B0 to F5 \citep{1998A&A...331..211M}. The HAeBe stars identified in \citetalias{2021RAA....21..288S} only include stars of spectral types B, and A. Hence, for completeness, we took spectra belonging to F0--F5 spectral type, as classified by LASP, and followed a similar ELS identification procedure as in \citetalias{2021RAA....21..288S}. This increased the total ELS sample size to 3850. Photometric magnitudes from 2MASS \citep{2003yCat.2246....0C}, ALLWISE \citep{2014yCat.2328....0C}, and Gaia EDR3 \citep{2020yCat.1350....0G}, and Gaia distance estimates from \cite{2021yCat.1352....0B} are obtained to estimate the stellar parameters of 3850 ELS. 

The IR excess in HAeBe stars suggests the presence of hot and/or cool dust in the circumstellar medium \citep{1992ApJ...397..613H, 1998ARA&A..36..233W}, whereas in the case of Classical Be (CBe) or Classical Ae (CAe; \citealp{2021MNRAS.501.5927A}) stars, the IR excess is attributed to thermal \textit{bremsstrahlung} emission \citep{1974ApJ...191..675G}. In general, IR excess in HAeBe stars is relatively higher than CBe/CAe stars. \cite{1984A&AS...55..109F} suggested a method to distinguish HAeBe stars from CBe/CAe stars based on the (\textit{H}- \textit{K$_s$})$_0$ color. The \textit{J}, \textit{H}, \textit{K$_s$} magnitudes (we included only those with $ph\_qual=AAA$) for the sample of 3850 stars from the 2MASS point source catalog \citep{2003yCat.2246....0C} were corrected for extinction (A$_V$). The \textit{A$_V$} values are retrieved from the 3D dust map of \cite{2019ApJ...887...93G}. A color-color diagram (CCDm) of the ELS is plotted between the de-reddened colors, (\textit{J}-\textit{H})$_0$ and (\textit{H}-\textit{K$_s$})$_0$, in Figure \ref{fig:CCD}. By employing the color criterion of (\textit{H}-\textit{K$_s$})$_0$ > 0.4 mag \citep{1984A&AS...55..109F} to the sample of ELS, we found that 127 are candidate HAeBe stars. They are located towards the right of the reddening band in Figure \ref{fig:CCD}. Moreover, the excess from the hot/cool dust present in the disc of YSOs is quantified by calculating the spectral indices/Lada indices, based on the steepness of the slope in the IR region in the SED \citep{1987IAUS..115....1L}. We estimated the spectral index, \textit{$n_{(K_{s}-W2)}$}, with 2MASS \textit{K$_s$} and WISE \textit{W2} magnitudes. Objects having \textit{$n_{(K_{s}-W2)}$} > -1.5 were added into the HAeBe category \citep{1993ApJ...406..122A, 2019yCat..51570159A, 2021MNRAS.501.5927A, 2021RAA....21..288S}, which increased the sample size by 24. Hence, by combining the H$\alpha$ emission information with the near- and mid-infrared photometric criteria we identified a sample of 151 HAeBe stars, which will be used for further analysis.

From the visual check of the 151 LAMOST spectra, we noticed that 32 out of 151 stars exhibit either one or both of the [O{\sc iii}] 4959 \AA, and 5007 \AA~forbidden emission lines. These emission lines are generally observed in nebular sources \citep{2017PASP..129h2001P}. We removed 32 stars showing these forbidden lines from our sample. Hence, the sample of HAeBe stars used for the present study is reduced to 119, after applying the photometric and spectroscopic constraints.

\section{Results and Discussion}
\label{sec:3}

In this section, we investigate the stellar properties of 119 HAeBe stars. We analyzed the spatial distribution of the HAeBe stars in the Galaxy. We did a spectral type distribution analysis of the sample of stars and compared them with the literature. A visual check on all the low-resolution spectra is done and the major spectral features observed in the HAeBe stars are discussed. We also explained the method to estimate extinction from Diffuse Interstellar Bands (DIB) for individual stars. Further, the correlation between the stellar parameters such as age, mass, and mass accretion rates are evaluated.

\subsection{Distribution of stars in the Galaxy}
\label{subsect:3.1}

\begin{figure}
	\includegraphics[width=1\columnwidth]{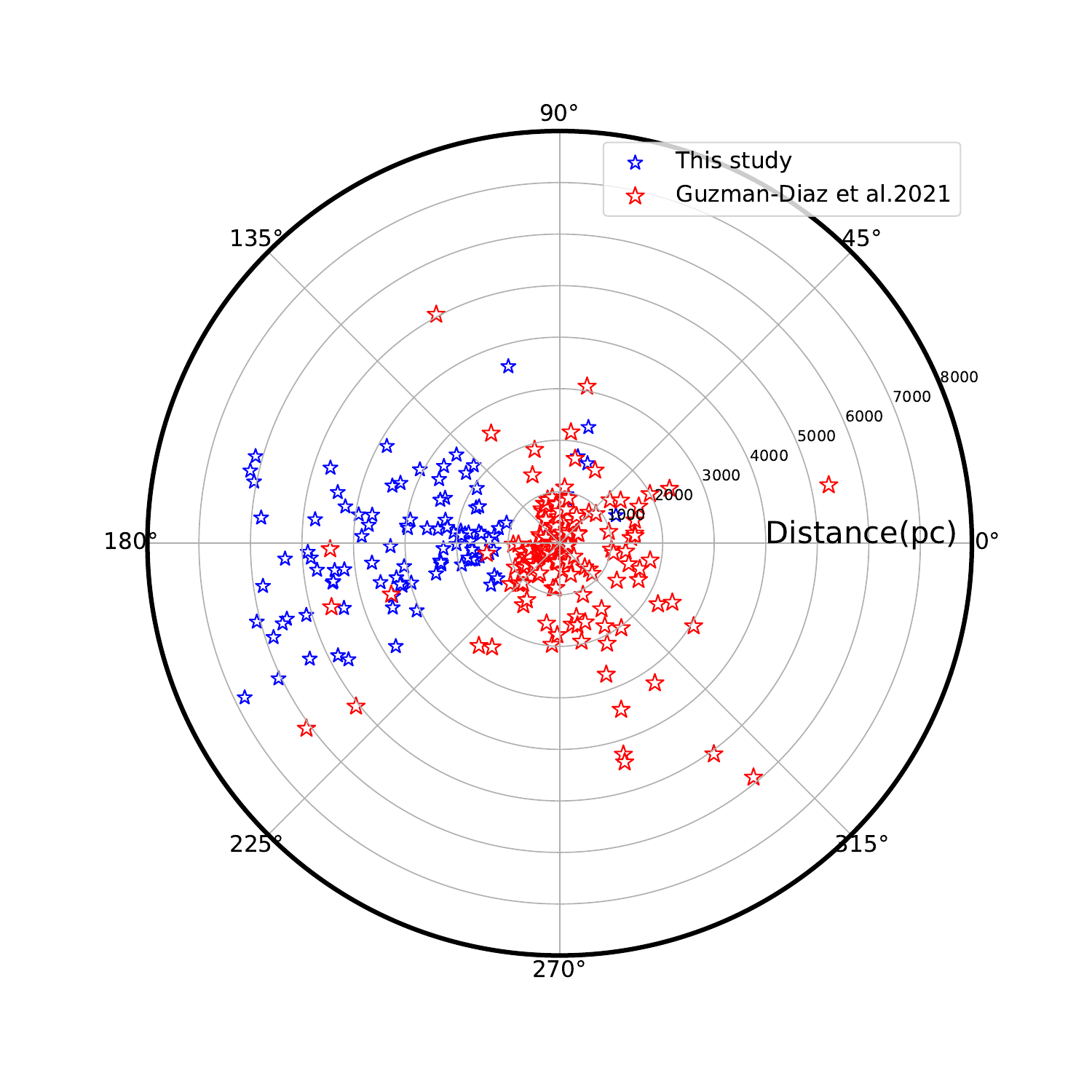}
    \caption{The distribution of 119 HAeBe stars from this study and 209 known HAeBe stars from \protect\citet{2021A&A...650A.182G} is shown in the Galactic Longitude (\textit{l}) vs Distance (\textit{d}) plane, in blue and red color, respectively. The distance values are adopted from \protect\citet{2021yCat.1352....0B}.} 
    \label{fig:Polar}
\end{figure}

The observational strategy of LAMOST Spectroscopic Survey of the Galactic anti-center (LSS--GAC) offers a unique opportunity to study the distribution of stars along the Galactic anti-center direction of the Milky Way \citep{2014IAUS..298..310L}. Initially, the LSS--GAC survey included a sky area of 3,400 $deg^2$, covering the Galactic longitudes $150\degree \leq l \leq 210\degree$ and latitudes $|b| \leq30\degree$ \citep{2015MNRAS.448..855Y}. Figure \ref{fig:Polar} represents the polar distribution of 119 HAeBe stars identified from LAMOST DR5. We have represented the position of the stars in the Galactic longitude (\textit{l}) vs distance (\textit{d}) plane, where the distance estimates of HAeBe stars used in this study are queried from \cite{2021yCat.1352....0B}. It can be seen from Figure \ref{fig:Polar} that 95\% HAeBe stars identified from this study are found to be within $135\degree \leq l \leq 225\degree$, i.e., towards the Galactic anti-center direction. On the other hand, majority of the well-studied HAeBe stars taken from \cite{2021A&A...650A.182G} are found to be homogeneously spread around the Sun. Hence, the LAMOST survey presents an opportunity to explore a new sample of HAeBe stars from a less-studied region of the Galaxy.

\subsection{Spectral type distribution}
\label{subsec:3.2}

\begin{figure}
	\includegraphics[width=\columnwidth]{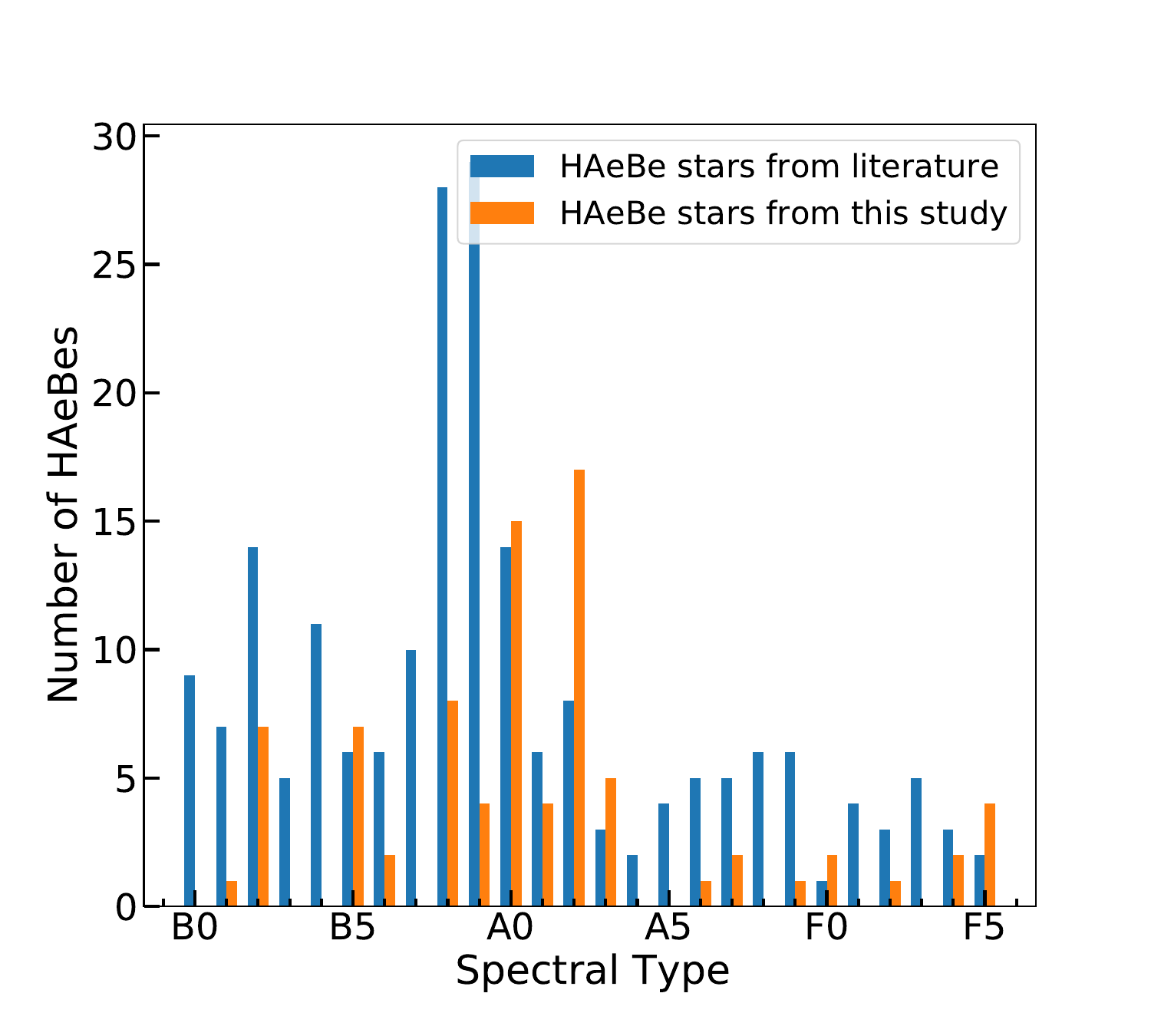}
    \caption{The histogram shows the comparison of the distribution of spectral types of 84 stars from the sample list in this study, in orange color, with 192 bona fide stars, in blue color, compile from \protect\citet{2021A&A...650A.182G}.}
    \label{fig:spec}
\end{figure}

One of the major criteria for the identification of HAeBe stars, as defined by \cite{1960ApJS....4..337H}, was the spectral type being A-type or earlier. The proposed criteria is too confining, leading to the rejection of probable objects of similar nature. Thus, the spectral type has been extended to include F-type stars, which showed emission lines in their spectra \citep{1998A&A...331..211M}. Discrepancies in the estimation of spectral type have been observed by employing different methods \citep{1972ApJ...173..353S, 1979ApJS...41..743C, 2001A&A...378..116M, 2004AJ....127.1682H}. Hence in this work, we have applied a consistent scheme for spectral re-classification of HAeBe stars.  

The spectral type of each LAMOST DR5 spectrum is estimated using LAMOST stellar parameter pipeline (LASP). A set of 183 stellar spectra from the pilot survey and LAMOST DR1 \citep{2014AJ....147..101W} serve as templates in their pipeline for classifying stars into each spectral type \citep{2019ASPC..523...91K}. Due to the lack of stars with certain spectral types, not all spectral type templates were available. So, it is quite possible that the stars which were cataloged as early A or F-type in LAMOST can be late B or A-type, respectively. Hence, the spectral type is re-estimated by a semi-automated template matching method using the MILES spectral library \citep{2006MNRAS.371..703S}. MILES includes the spectra of 980 stars covering a wavelength range of 3525 -– 7500 \AA, with a spectral resolution of 2.5 \AA~\citep{2011A&A...532A..95F}. We fitted the LAMOST spectra with 3 best-fitting templates generated using a python routine. As a preliminary check, we matched the H$\delta$ and H$\epsilon$ profiles with the templates. For B-type stars, we used the absorption strength of He{\sc i} 4471 \AA~and Mg{\sc ii} 4481 \AA. For A- and F-type stars, we used the absorption strength of Mg{\sc ii} 4481 \AA~and Ca{\sc ii} 3933 \AA~for spectral estimation. The template spectra that visually fits the best, were used to define the spectral type of HAeBe stars. It should be emphasized that one can anticipate an error of up to two sub-classes when calculating spectral type from low-resolution spectra. Spectral type for 35 stars was not assigned because the blue end of the spectra was noisy. Further in Figure \ref{fig:spec}, we compared the spectral type distribution in our sample with the bona fide sample of HAeBe stars compiled from \cite{2021A&A...650A.182G}. It can be seen from the figure that the histogram distribution of early HBe stars from the literature peaks at B2 -- B4, whereas for our sample it is B2 -- B5. Further, another peak is seen at B9 –- A1 for the literature stars whereas it is A0 –- A2 for our stars, with a decrease in number towards the late type.

\subsection{Brief overview of the spectral lines present in our sample of HAeBe stars}
\label{subsec:3.3}

\begin{figure*}
	\includegraphics[width=2.1\columnwidth]{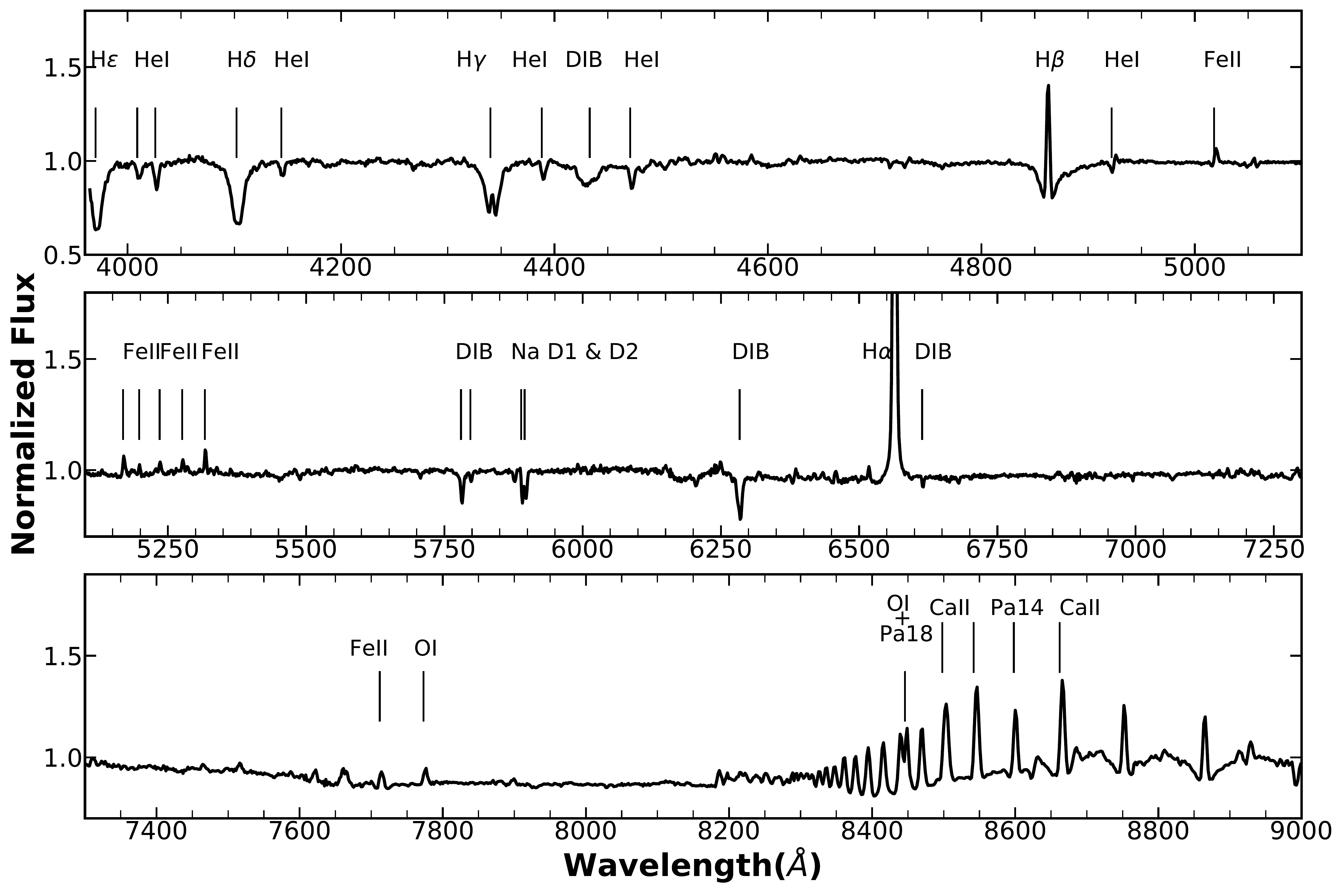}
    \caption{A representative spectrum of a HAeBe star from the present study with LAMOST ID J052019.20+320817.0 and spectral type B2IV. Prominent spectral features observed in the wavelength range of 3800 -- 9000 \AA~are labeled. H$\alpha$ -- 6563 \AA, H$\beta$ -- 4861 \AA, H$\gamma$ -- 4340 \AA, H$\delta$ -- 4102 \AA, H$\epsilon$ -- 3970 \AA; He{\sc i} -- 4009, 4026, 4144, 4387, 4471, 4992, 5876 \AA; DIB -- 4428, 5780, 5797, 6284, 6614 \AA; Fe{\sc ii} -- 5018, 5169, 5198, 5235, 5276, 5317, 7712 \AA; Na{\sc i} -- 5890, 5896 \AA; O{\sc i} -- 7773, 8446 \AA; Ca{\sc ii} -- 8498, 8542, 8662 \AA~(Ca{\sc ii} triplet lines are blended with P16, P15 and P13 emission lines, respectively). Also, Paschen series (P11 -- P21) are visible in emission.}
    \label{fig:spec_rep}
\end{figure*}

The spectrum of a HAeBe star is rich in emission lines that originate from the circumstellar disc. This section describes the spectral lines seen in the spectra of our sample of HAeBe stars. For this purpose, we visually checked the 119 HAeBe spectra identified from LAMOST DR5, which covers the spectral range of 3800 – 9000 \AA. Figure \ref{fig:spec_rep} shows a representative spectrum of a HAeBe star with the LAMOST ID J052019.20+320817.0.

\subsubsection{The H$\alpha$ line}
\label{subsubsec:3.3.1}

\begin{figure}
	\includegraphics[width=1\columnwidth]{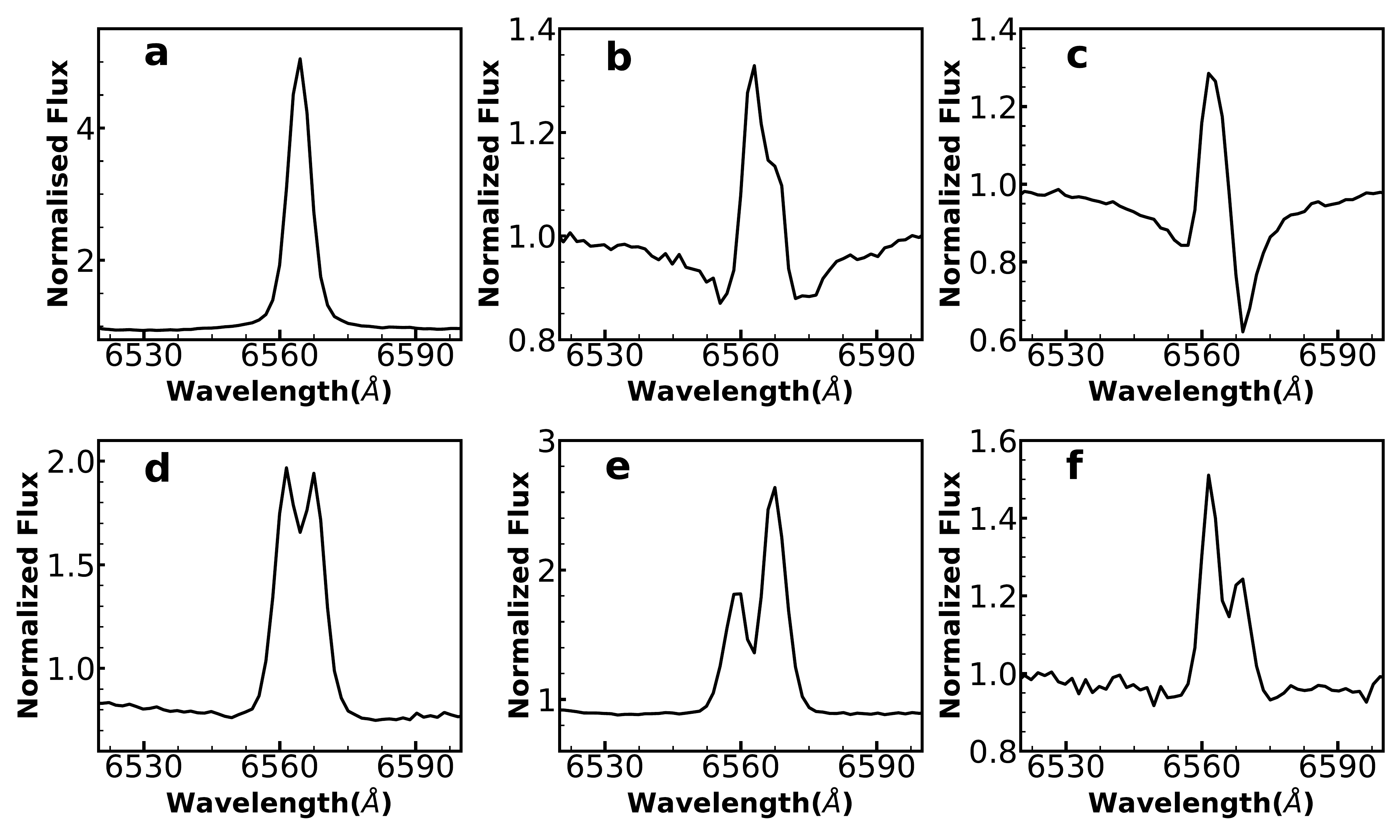}
    \caption{H$\alpha$ emission profiles observed in our sample. (a), (b), (c) shows single-peak emission, single-peak emission in absorption and asymmetric single-peak emission, respectively, (d) represents the double-peak emission, (e) and (f) shows the double-peak emission with asymmetry in its violet and red components.}
    \label{fig:morp}
\end{figure}

The presence of Balmer series in emission, particularly H$\alpha$ at 6563 \AA~categorizes a star into ELS, and it indicates the presence of a circumstellar disc/environment \citep{1974ApJ...191..675G, 1992ApJS...82..285H}. HAeBe stars in our sample present various emission line profiles of H$\alpha$. \cite{1984A&AS...55..109F} proposed an H$\alpha$ line profile classification system for the HAeBe stars. A similar approach is taken in this study based on the visual check where we classified the profiles into emission with a single-peak (sp), emission with a single-peak in absorption (eia), and emission with a double-peak (dp). Figure \ref{fig:morp} shows the different H$\alpha$ profiles observed in our sample. Single-peak H$\alpha$ profile (Figure \ref{fig:morp}a) is observed for 43.7\% stars, and 35.3\% show emission with a single-peak in absorption (Figure \ref{fig:morp}b). Asymmetry in H$\alpha$ emission is observed in nine stars, where an emission line is formed inside an absorption trough (Figure \ref{fig:morp}c). Similar H$\alpha$ profiles showing asymmetry are reported for eight bonafide HAeBe stars from the literature. Double-peak profiles are observed in 11 stars (Figure \ref{fig:morp}d). In the case of the double-peak profile, we observed that in some spectra the violet (V) and red (R) components of the double-peak emission profile show different emission strengths (Figure \ref{fig:morp}e and Figure \ref{fig:morp}f). The double-peak profile is due to the emission seen from the disc at a large inclination angle, which is commonly observed in CBe stars \citep{1992ApJS...81..335S}. The double-peaked profiles may not be completely resolved in a few cases due to the low spectral resolution (R$=$1800) of the LAMOST spectrum. The H$\alpha$ profile of 14 spectra could not be classified in any of the above categories due to high noise.

To quantify the extent of the circumstellar disc, we computed the equivalent width of H$\alpha$ emission line, denoted as EW(H$\alpha$). The EW(H$\alpha$) of HAeBe stars is measured using the tasks in Image Reduction and Analysis Facility (IRAF; \citealp{1986SPIE..627..733T}). For ELS, the H$\alpha$ appears in emission only after filling the underlying absorption profile. Hence, the photospheric absorption component of all the HAeBe stars has to be accounted for while estimating the net emission strength. The absorption strength of H$\alpha$ is measured from the MILES stellar library of similar spectral type as that of the HAeBe star. This is further added to the emission component to estimate the net H$\alpha$ EW.

The corrected EW(H$\alpha$) measurements of the known HAeBe stars from \cite{2018A&A...620A.128V} were compiled. It is then compared with our sample in Figure \ref{fig:EW}, where the distribution of corrected EW(H$\alpha$) against the spectral type is represented. From the measurements obtained, 80\% of the stars in our sample have EW(H$\alpha$) from -4 \AA~to -60 \AA, which is in agreement with the EW(H$\alpha$) distribution of HAeBe stars reported in the literature \citep{2018A&A...620A.128V}. Here, we observe that the EW(H$\alpha$) is higher for HBe stars, which subsequently decreases for late-type stars. A similar trend is observed for EW(H$\alpha$) of HAeBe stars from the literature. Such variation in the EW(H$\alpha$) in HAeBe stars may be caused by the different contributors of the H$\alpha$ profile over the range of HAeBe stars. In other words, the major H$\alpha$ contributors in HBe stars are from the disc and wind while for the HAe stars, it may result from disc wind, and the accretion column due to magnetospheric accretion \citep{2017MNRAS.464.1984M}.

\subsubsection{Fe{\sc ii} lines}
\label{subsubsec:3.3.2}

Various Fe{\sc ii} lines belonging to different multiplet series have been detected in HAeBe stars \citep{1982ESASP.176...99T, 1992ApJS...82..285H}. We identified 22 different Fe{\sc ii} lines belonging to seven multiplets across 77 (65\%) stars. Fe{\sc ii} lines in 15 stars are not distinguishable due to the low SNR in the spectra. The most common Fe{\sc ii} lines observed are 5018 \AA~and 5169 \AA~(multiplet 42), which is in agreement with the observation made by \cite{2004AJ....127.1682H}. Other Fe{\sc ii} lines such as 4584 \AA~(multiplet 38), 5198, 5235, 5276 \AA~(49), 5317 \AA~(48, 49) and 7712 \AA~(73) are commonly observed in our HAeBe stellar spectra. In addition, a few stars also show 4233, 4417, 4303 \AA~(27), and 6516 \AA~(40) Fe{\sc ii} lines. 

The distribution of EW of Fe{\sc ii} 5169 \AA~line with respect to spectral type is shown in Figure \ref{fig:EW}. Similar to H$\alpha$, other lines in ELS appear in emission only after filling the underlying photospheric absorption profile. Hence the absorption correction has to be done to calculate the effective Fe{\sc ii} emission EW for a HAeBe star. We measured the absorption strength at Fe{\sc ii} from the MILES stellar library \citep{2006MNRAS.371..703S}, corresponding to the spectral type of each star. This absorption component is added to the measured emission EW to estimate the corrected Fe{\sc ii} EW. Similarly, the EW of Fe{\sc ii} 5169 \AA~line compiled from the literature is also corrected. The corrected EW of Fe{\sc ii} 5169 Å line for our sample of 40 HAeBe stars and those compiled from \cite{2004AJ....127.1682H} for 15 known HAeBe stars are represented in Figure \ref{fig:EW}. The EW of Fe{\sc ii} is found to be in the range 0 to -2 \AA. 

\cite{1984A&A...140...49F} suggested that the excitation of Fe{\sc ii} ions by L$\gamma$ and the higher-order members of the Lyman series (other than L$\alpha$ and L$\beta$), along with the contribution from the Lyman continuum, can result in the formation of Fe{\sc ii} emission lines. The presence of Balmer emission ensures the presence of Lyman emission. In such a scenario, they suggested an autoionization mechanism, where the ground state Fe{\sc i} atom is ionized by a Lyman line photon, which also simultaneously excites the Fe{\sc ii} ion to the upper level of the observed emission transition. The subsequent downward cascading produces the Fe{\sc ii} emission lines. This suggests that the autoionization mechanism might be responsible for the formation of Fe{\sc ii} emission lines in HAeBe stars. They probed the viability of their prediction by comparing the Fe{\sc ii} multiplets formed through autoionization mechanism with those seen in the spectra of stars. We also compared the Fe{\sc ii} multiplets in emission from our sample with \cite{1984A&A...140...49F}. Multiplet 27, 38, 40, 42, 48, and 49 are observed, both in our sample and from the literature. This gives us a stronger lead to investigate the formation mechanism of Fe{\sc ii} lines in YSOs in future works.

\subsubsection{He{\sc i} lines}
\label{subsubsec:3.3.3}

\begin{figure}
	\includegraphics[width=\columnwidth]{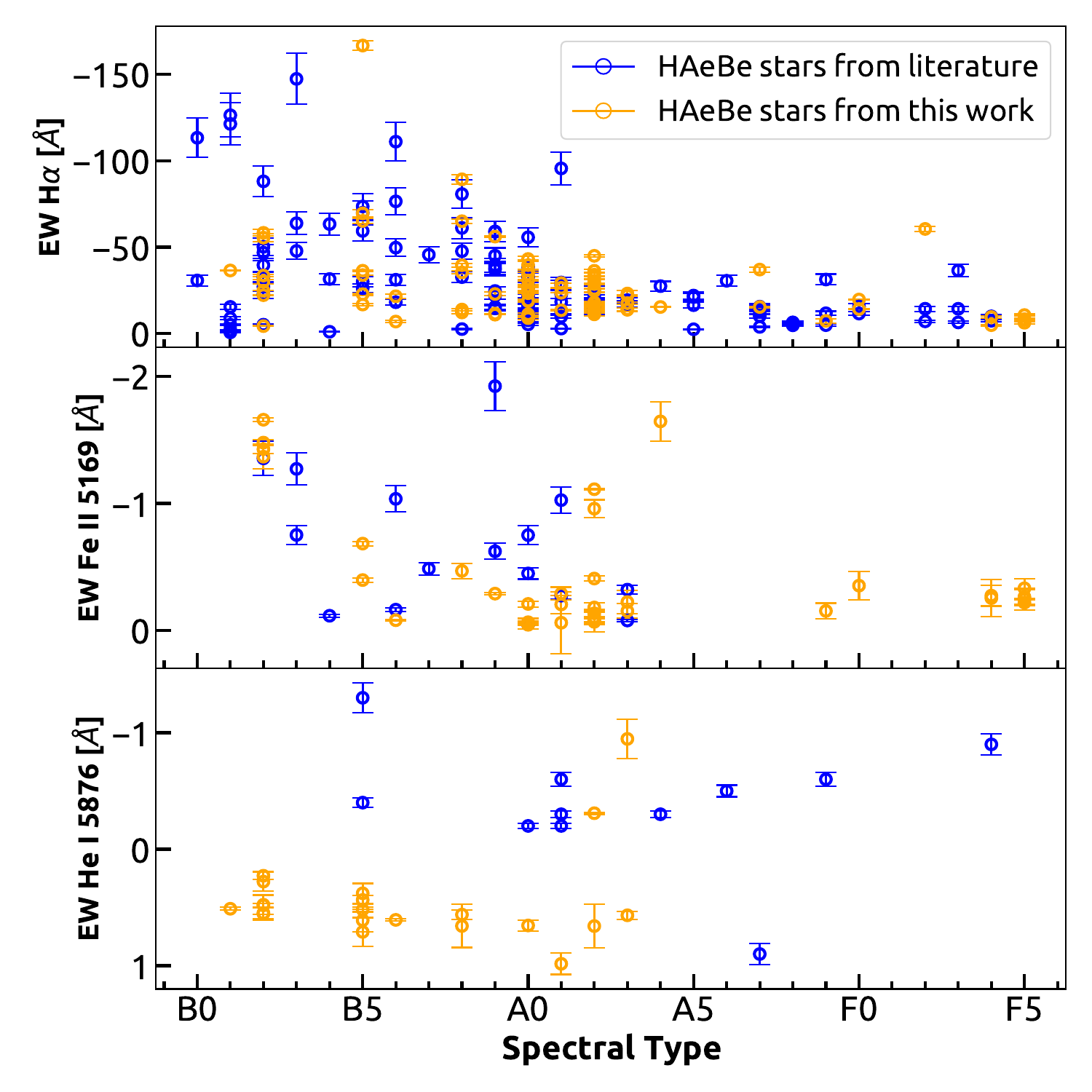}
    \caption{The top, middle, and bottom panels represent the distribution of EW of H$\alpha$, Fe{\sc ii} 5169 \AA~and He{\sc i} 5876 \AA~lines observed in our sample of HAeBe stars, respectively, against the spectral type. The blue circle represent the HAeBe stars from literature and the orange circle represents the HAeBe stars from this study. The EW(H$\alpha$) for stars from the literature is taken from \protect\citet{2018A&A...620A.128V}, EW of Fe{\sc ii} 5169 \AA~and He{\sc i} 5876 \AA~is adopted from \protect\citet{2004AJ....127.1682H}. }
    \label{fig:EW}
\end{figure}

\begin{figure}
	\includegraphics[width=\columnwidth]{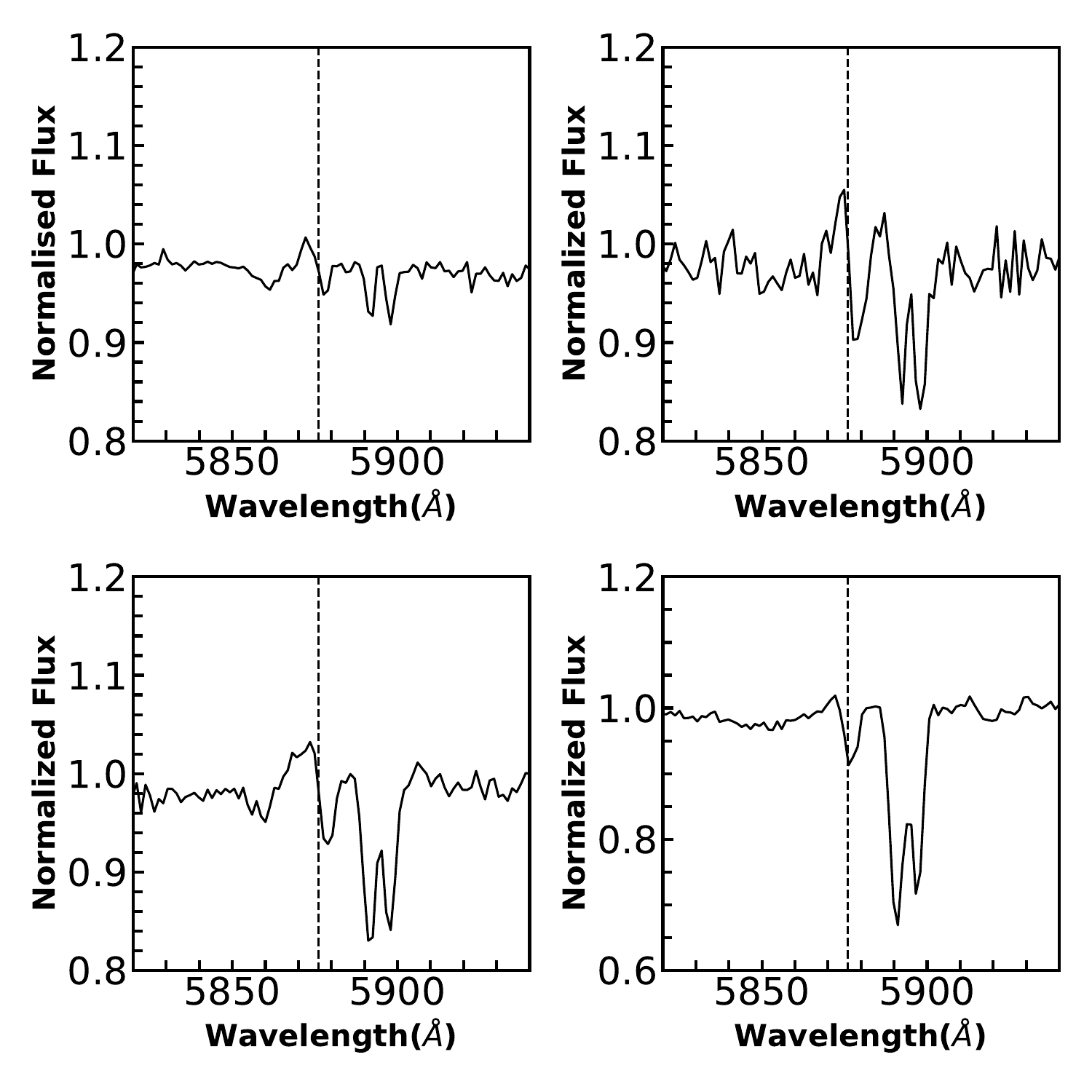}
    \caption{The figure shows the inverse p-cygni profile of He{\sc i} 5876 \AA~line for 4 HAeBe stars.}
    \label{fig:He}
\end{figure}

The neutral and ionized Helium lines (He{\sc i} and He{\sc ii}, respectively) appear only in hot stars and are absent in the spectrum of stars cooler than A0 \citep{1993ASPC...40..665B, 1993A&A...278..187C}. Also, He{\sc i} lines have been studied as wind and activity tracers in HAeBe stars in the literature. In the present study, He{\sc i} lines are observed in 72 stars (53.73\%). The most common He{\sc i} line observed is at 5876 \AA. The presence of redshifted He{\sc i} 5876 \AA~absorption line in HAeBe stars is considered to be an evidence of infall by \cite{1992ApJS...82..285H}. Also, \cite{2004AJ....127.1682H} mentioned that He{\sc i} 5876 \AA~line can be affected by absorption external to the photosphere. Apart from these, studies of individual stars by \cite{1995A&A...301..155B} showed that the stars with He{\sc i} 5876 \AA~profile, as shown in Figure \ref{fig:He}, does not have a photospheric origin. Such profile, namely inverse P-Cygni, is composed of two components: an absorption component centered on the photospheric rest wavelength formed in a spherical slab surrounding the stellar photosphere and a red-shifted absorption component presumably influenced by the boundary layer accretion. Figure \ref{fig:He} represents a similar profile of He{\sc i} 5876 \AA~observed in four HAeBe stars from this study. Although the HAeBe stellar spectra show emission lines, He{\sc i} absorption lines are also observed at 4009, 4026, 4144, 4387, 4471, 4713, 4922, 5016, 5876, 6678 and 7065 \AA. Only two stars show He{\sc i} 5876 \AA~in emission; J055931.30+201959.8, and J052135.30-045329.2. The measured equivalent width of He{\sc i} 5876 \AA~line in 21 HAeBe stars from this study and equivalent width compiled from the \cite{2004AJ....127.1682H} for 11 known HAeBe stars are represented in Figure \ref{fig:EW}. Photospheric absorption correction for He{\sc i} 5876 \AA~is not performed because the respective absorption core in the template spectrum is not measurable. From Figure \ref{fig:EW} it can be seen that, for our sample of stars, He{\sc i} emission ceases to appear for stars later than A3. Interestingly, among the sample of HAeBe stars from the literature, He{\sc i} emission is seen for later spectral types. This dichotomy will be explored in detail in a future work.

\subsubsection{Ca{\sc ii} triplet lines}
\label{subsubsec:3.3.4}

Ca{\sc ii} triplet (8498, 8542, 8662 \AA) observed in the HAeBe stellar spectra are non-photospheric lines. Ca{\sc ii} triplet lines are also studied as activity tracers in HAeBe stars. \cite{1992ApJS...82..285H} studied the Ca{\sc ii} triplet emission present in 27 HAeBe stars. They found a higher detection rate of Ca{\sc ii} lines in HAeBe than CBe stars because Ca{\sc ii} requires a more shielded and cooler environment. They also noticed that the surface flux of Ca{\sc ii} 8542 \AA~is increasing with the effective temperature. The same trend was verified by \cite{1995A&A...301..155B}, who interpreted the correlation to the presence of an activity in HAeBe stars, possibly originating in the boundary layer between accretion disc and the stellar surface.

In this study, we found 61 (45.5\%) stars showing Ca{\sc ii} triplet lines. Among these, a subset of 37 stars show Ca{\sc ii} triplet in emission while 24 stars show it in absorption. All the B-type stars (30 out of 37 stars) from our sample are showing Ca{\sc ii} triplet in emission. Ca{\sc ii} triplet is found to be blended with the Paschen lines P16 (8502 \AA), P15 (8545 \AA) and P13 (8665 \AA), respectively. We used the method discussed in \cite{2021MNRAS.500.3926B} to remove the Paschen component from the Ca{\sc ii} lines, which is as follows. We measured the equivalent width of each of the Ca{\sc ii} triplet lines (which is blended with the adjacent Paschen line emission) and subtracted the equivalent width of P14 (8598 \AA), which is very similar to that of the Paschen lines P13, P15 and P16, to retrieve the EW of Ca{\sc ii} triplet lines. The deblended EW of Ca{\sc ii} triplets are given in table \ref{tab:A1}.

\subsubsection{Diffuse Interstellar Bands}
\label{subsubsec:3.3.5}

The DIB are a set of absorption features with broader line widths than the stellar absorption lines, which are observed in the optical-infrared wavelength region \citep{2015ApJ...798...35Z, 2019AcA....69..159K}. These absorption bands were identified due to interstellar origin by \cite{1934PASP...46..206M} and \cite{1938ApJ....87....9M}, followed by vast research on the plausible candidates of DIB carriers. \cite{1975ApJ...196..129H} listed 39 DIB from 4400 \AA~to 6850 \AA. Through the advent of better instrumentation and advancement in spectroscopy, more than 500 DIB have been identified till now \citep{2008ApJ...680.1256H, 2011Natur.479..200G, 2014A&A...569A.117C, 2016ApJ...821...42H, 2019ApJ...878..151F}. In this study, we report five DIB observed at 4428 \AA, 5780 \AA, 5797 \AA, 6284 \AA~and 6614 \AA. It is to be noted that 50\% of the sample shows both 5780 \AA~and 6284 \AA~in absorption.

\subsubsection{Other Spectral features}
\label{subsubsec:3.3.6}

We identified five different forbidden emission lines, that are generally observed in HAeBe stars, across 66 stars from our sample. The most commonly observed forbidden lines are [{O\sc i}] 5577 \AA~(57 stars), 6300 \AA~(27 stars), 6363 \AA~(4 stars) and [{S\sc ii}] 6717\AA~, 6731 \AA~(29 stars). The blue-shifted [O{\sc i}] 6300 \AA~are formed from the low-density and collisionally excited gas \citep{1985A&A...151..340F, 2003AJ....126.2971V}, indicating that these lines are formed in the outflow from the stars \citep{1997A&A...321..189C}. Among the spectra showing [O{\sc i}] and [S{\sc ii}], 20 stars display detectable levels of [N{\sc ii}] 6548, 6584 \AA~forbidden doublet in emission, which is predominantly detected in those stars with jets or nebulosity. These lines indicate low-density gas envelopes or stellar winds associated with the PMS stars  \citep{1994ApJS...93..485H}. Transitions of [Fe{\sc iii}] 4659 \AA~and [N{\sc i}] 5200 \AA~are also detected in one star each. 

HAeBe stars in our sample also exhibit emission or absorption spectral features other than the prominent lines mentioned above. We found Ca{\sc i} 4226 \AA~in absorption in nine HAeBe stars, with spectral type later than A9. Usually, they are seen in emission in T Tauri stars \citep{1945ApJ...102..168J}. Also, in the spectra of a few HAeBe stars, the transitions of Mg{\sc ii} 4481 \AA~and Si{\sc ii} 6347 \AA, 6371 \AA~are seen. The strength of Mg{\sc ii}  absorption line increases towards late-type HAeBe stars. Si{\sc ii} lines are observed in stars earlier than A2. Also, Sr{\sc ii} 4078 \AA~is found in four HAeBe stars of F spectral type, which are commonly observed in late-type stars. Metallic blends at 6497 \AA~is also noted in a few HAeBe stars.

\subsection{Correlation analysis between the emission strengths Fe{\sc ii} 5169 \AA~and H$\alpha$}
\label{subsec:3.4}

\cite{1927ApJ....65..286M} noted that all stars showing Fe{\sc ii} emission lines also show H$\alpha$ emission line. \cite{1970ApJ...161L.105G} found a strong correlation between the presence of Fe{\sc ii} emission lines in eight stars with the IR excess from the circumstellar envelope. Studies on HAeBe stars also mentioned about a correlation between the emission strength of Fe{\sc ii} lines with H$\alpha$ and He{\sc i} lines \citep{1979ApJS...41..743C, 2004AJ....127.1682H}. We noticed that only a few studies have been carried out to understand Fe{\sc ii} emission in HAeBe stars. \cite{2004AJ....127.1682H} have found a correlation of 0.74 between the EW of Fe{\sc ii} 5169 [EW(Fe{\sc ii} 5169)] and EW(H$\alpha$). They categorized the Fe{\sc ii} emission and the anomalous absorption lines as non-photospheric features and are thought to be emitted from outside the stellar photosphere. 

\begin{figure}
	\includegraphics[width=\columnwidth]{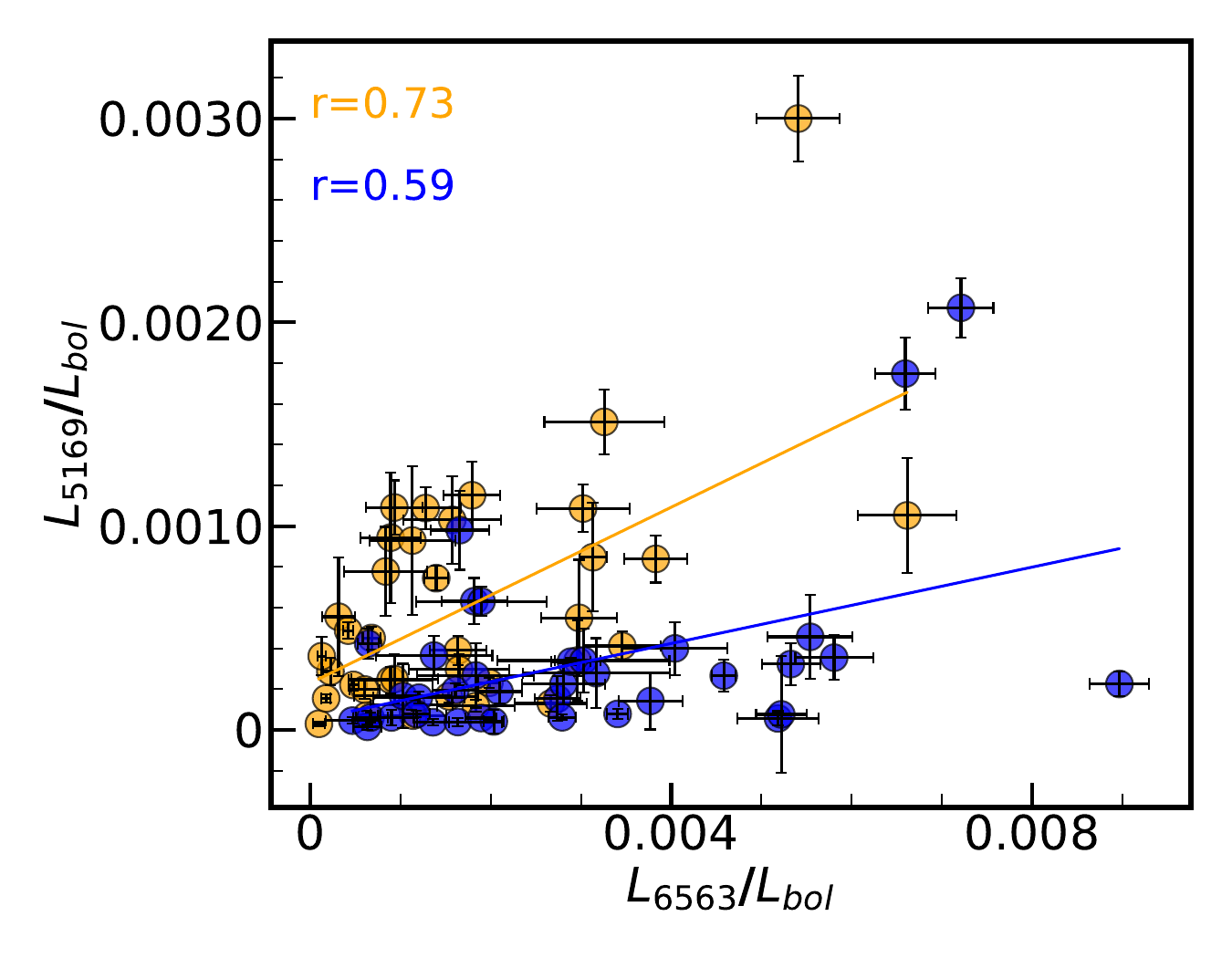}
    \caption{The $L_{line}/L_{bol}$ of H$\alpha$ is plotted against Fe{\sc ii} 5169 \AA~lines. The stars in orange color represent the $L_{line}/L_{bol}$ derived from the observed EW. The stars in blue color represent the $L_{line}/L_{bol}$ derived from the photospherically corrected EW. The best fit for both cases, before and after correction, is represented in orange and blue lines, respectively. }
    \label{fig:Feii}
\end{figure}

For the present study, we included 40 HAeBe stars which show both Fe{\sc ii} and H$\alpha$ lines in emission. Figure \ref{fig:Feii} represents the correlation of line to stellar luminosity ratio ($L_{line}/L_{bol}$) between Fe{\sc ii} 5169 \AA~and H$\alpha$. The correlation coefficient is 0.73. We also re-checked the correlation, once the absorption correction was done (see sections \ref{subsubsec:3.3.1} and \ref{subsubsec:3.3.2}). The correlation after absorption correction is reduced to 0.59. The correlation coefficient in both cases, before and after the absorption correction, is moderate. Such a weak correlation implies that Fe{\sc ii} emission might not be following H$\alpha$, but rather may be formed in a distinct region in the inner gaseous disc. Various studies in the literature have shown that H$\alpha$ emission strength in HAeBe stars is due to the contribution from disc, wind, and accretion column \citep{2006MNRAS.370..580K, 2017MNRAS.464.1984M}. However, \cite{1993A&A...278..187C} studied the Fe{\sc ii} 5018 \AA~lines appearing in the star AB Aur, and suggested that the Fe{\sc ii} 5018 \AA~might form from the bulk of the expanding chromosphere.

Interestingly, the emission-line regions of H$\alpha$ and Fe{\sc ii} are evaluated in the context of CBe stars, which are close analogs to HAeBe stars. In CBe stars, \cite{1992ApJS...81..335S} found H$\alpha$ emission arises in the range 7 -- 19 stellar radii, while the Fe{\sc ii} emission is in 3 -- 4 stellar radii from the central star. Similarly, \cite{2006A&A...460..821A} also suggested that the formation region of Fe{\sc ii} emission line in the circumstellar disc cannot be situated far from the central star and derived an average radius of, $R/R_{*}$ = 2.0 $\pm$ 0.8. It is evident from these studies that the Fe{\sc ii} emission in HAeBe stars may be similar to that of CBe -- closer to the star and inside the H$\alpha$ emission region in the disc. A moderate correlation between the emission strengths of H$\alpha$ and Fe{\sc ii} 5169 \AA~suggests a possible common emission region for Fe{\sc ii} lines and one of the components of H$\alpha$.

In CBe stars, Fe{\sc ii} emission is from a region in the disc closer to the star than from where H$\alpha$ emission is happening. A similar study by \cite{1987A&A...173..299H}, \cite{1992ApJS...81..335S}, and \cite{2021MNRAS.500.3926B} found no correlation between the EW of Fe{\sc ii} 7712 \AA~and EW(H$\alpha$) in their sample of CBe stars. No such studies on Fe{\sc ii} 7712 \AA~have been reported for HAeBe stars. Hence, we investigated Fe{\sc ii} 7712 \AA~and H$\alpha$ emission lines for any possible correlation between the EW. No correlation has been found, implying the region of formation of Fe{\sc ii} lines may vary for each multiplet. Hence, it is evident from the present analysis that, a) Fe{\sc ii} emission line is due to the contribution from various emission regions around the star, b) One of the Fe{\sc ii} emission region could be in the inner disc, with an overlap with one component of the H$\alpha$ emitting region, due to which we find a moderate correlation between the emission strengths, and c) since we do not see any correlation between Fe{\sc ii} 7712 and H$\alpha$, it is quite possible that Fe{\sc ii} lines belonging to various multiplets may be forming at different regions around the star.

\subsection{Estimation of A$_V$ using DIB}
\label{subsec:3.5}

\begin{figure}
	
	\includegraphics[width=\columnwidth]{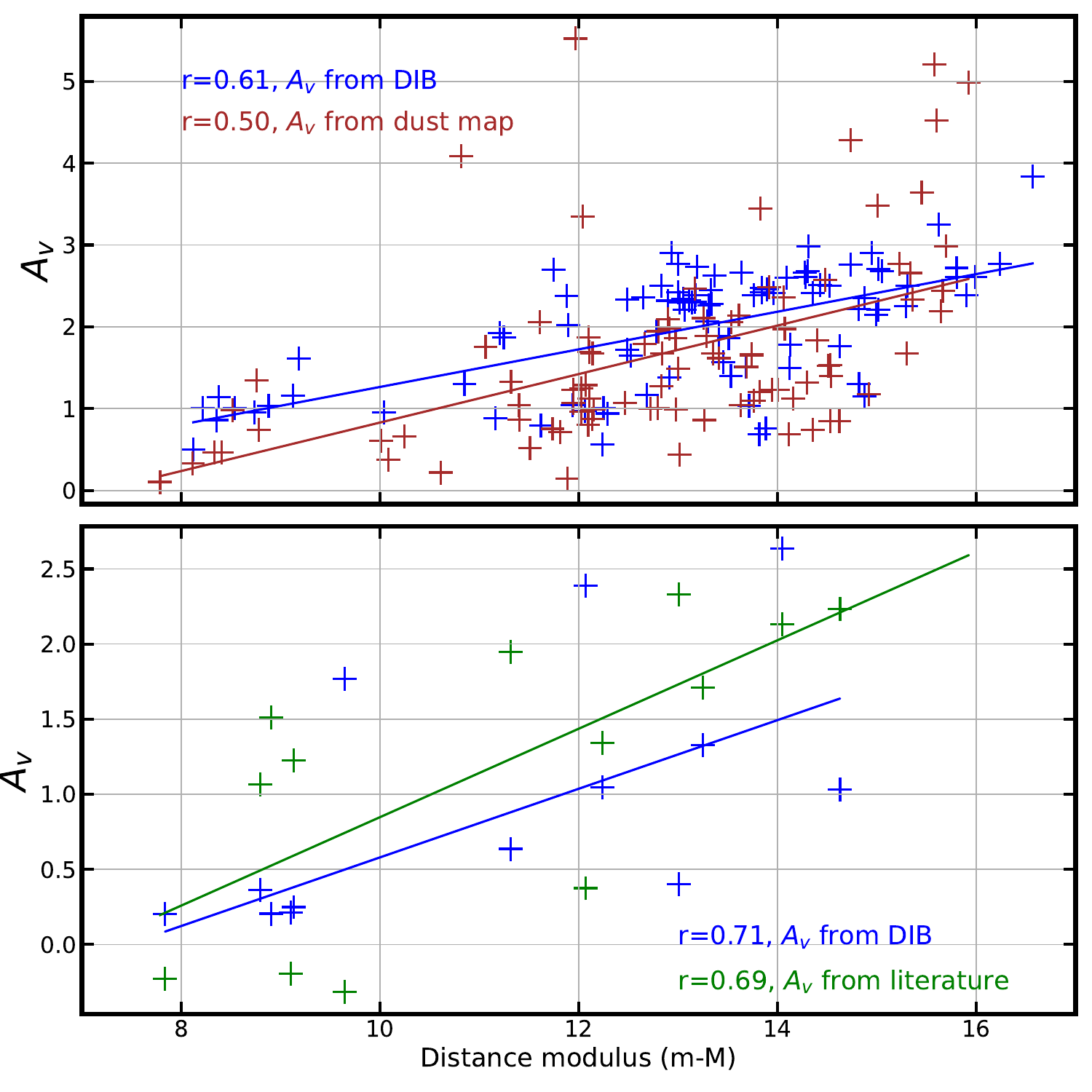}
    \caption{The plot compares the A$_{VD}$, A$_{VG}$, and A$_{VN}$ correlations against the distance modulus. In both panels, the A$_{VD}$ is represented by the blue color. In the top panel, A$_{VG}$ is represented in brown color and in the bottom panel, the A$_{VN}$ is represented in green color for the stars. The best fit is represented in respective colors.
}
    \label{fig:Av}
\end{figure}

The reddening due to interstellar material along the line-of-sight in the HAeBe stars cannot be neglected \citep{2006Ap&SS.305...11Z}. To calculate the extinction, we cross-matched the 119 HAeBe stars with NOMAD and APASS catalog where very few stars had the \textit{B} and \textit{V} magnitudes. The unavailability of enough data to estimate extinction motivated us to explore other techniques. Therefore in this paper, the magnitudes have been corrected for extinction using the interstellar spectral features, namely DIB (see section \ref{subsubsec:3.3.5}). While most of the studies have focused on identifying the responsible DIB carriers \citep{1998ApJ...493..217K, 2016JPhCS.728f2005G, 2021ECS.....5..381O}, in this study, we use them as a tool to estimate the reddening of the stars based on the correlation of DIB with the color excess, E(B-V). In literature, studies have established the correlation between E(B-V) and equivalent width of DIB ($EW_{DIB}$) measured from the stars \citep{1975ApJ...196..129H, 2003A&A...397..927P, 2008A&A...488..969M, 2012A&A...544A.136R}. Recently, \cite{2019AcA....69..159K} studied several DIB within the range of 5700 - 8700 \AA~and checked their correlations with E(B-V). One of the advantages of measuring the reddening through this method is that for a broad range of extinction, the correlations between different DIB and E(B-V) looks tight, which is caused most likely due to the presence of clouds along any line-of-sight \citep{2019AcA....69..159K}. 

In this study, we measured the $EW_{DIB}$ for 5780 \AA, 5797 \AA, 6284 \AA~and 6614 \AA~for the calculation of E(B-V). Since most of the HAeBe stars in our sample are not studied in detail, their respective E(B-V) values are not available in the literature. Thus, we used equation \ref{equ:1} given by \cite{2011ApJ...727...33F}, to estimate E(B-V) values for individual stars, as given below.
\begin{equation}
\label{equ:1}
E(B-V) = a + b * EW_{DIB}
\end{equation}

where the values of a and b coefficients are given in Table \ref{tab:coefficient}. The coefficients are adopted from \cite{2011ApJ...727...33F}, who used it to study the correlation of 8 DIB with E(B-V) for reddened stars. With this information, we calculated the visual extinction A$_V$  for 93 stars, using the total-to-selective extinction, R$_V$ (A$_V$ = R$_V$ * E(B-V)) value of 3.1. For the remaining stars with no DIB features, A$_V$ values are estimated using the extinction map of \cite{2019ApJ...887...93G}. The estimated A$_V$ values for all the HAeBe stars are used to extinction correct the Gaia magnitudes, \textit{G}, \textit{$G_{BP}$} and \textit{$G_{RP}$}, using the extinction curve of \cite{1990ARA&A..28...37M}, following the method explained in \cite{2019yCat..51570159A}. It should also be noted that no new objects were added or removed from the location of HAeBe stars from Figure \ref{fig:CCD}, after the correction of magnitudes/colors using the revised extinction/reddening estimates.

\begin{table}
\centering
\caption{DIB Coefficients for equation \ref{equ:1}}
\label{tab:coefficient}
\begin{tabular}{ccc}
\hline
DIB  & a                      & b                     \\
\hline 
5780 \AA & $(-8.36 \pm 3.84) \times 10^{-3}$  & $(1.98 \pm 0.01) \times 10^{−3}$ \\
5797 \AA & $(−2.86 \pm 0.57) \times 10^{-2}$  & $(5.74 \pm 0.06) \times 10^{−3}$ \\
6284 \AA & $(−7.71 \pm 0.78) \times 10^{−2}$ & $(9.57 \pm 0.17) \times 10^{−4}$ \\
6614 \AA & $(1.96 \pm 0.37) \times 10^{−2}$  & $(4.63 \pm 0.04) \times 10^{−3}$ \\
\hline
\end{tabular}
\end{table}

In order to validate that the DIB-based extinction values (A$_{VD}$) are more reliable for this study, we compared the median of the A$_{VD}$ values derived from the four DIB strengths with the A$_{VG}$ (A$_V$ from Green’s dust map) and A$_V$ values from the NOMAD (A$_{VN}$). From NOMAD, only 13 stars have B and V magnitudes. The A$_V$ estimated from the different methods is plotted against the distance modulus in Figure \ref{fig:Av}. A similar approach is observed in Mendigutia et al. (1999), where reddening is plotted against the distance modulus. Beyond a higher distance modulus value of 10 -– 11, an extinction/reddening spread is seen in both cases. It is evident that A$_{VD}$ has a higher correlation with the distance modulus than A$_{VG}$. Hence, we demonstrated that A$_{VD}$ values are more reliable indicators of extinction for our sample.

\subsection{Estimation of Age and Mass of HAeBe stars}
\label{subsec:3.6}

\begin{figure}
	
	\includegraphics[width=1\columnwidth]{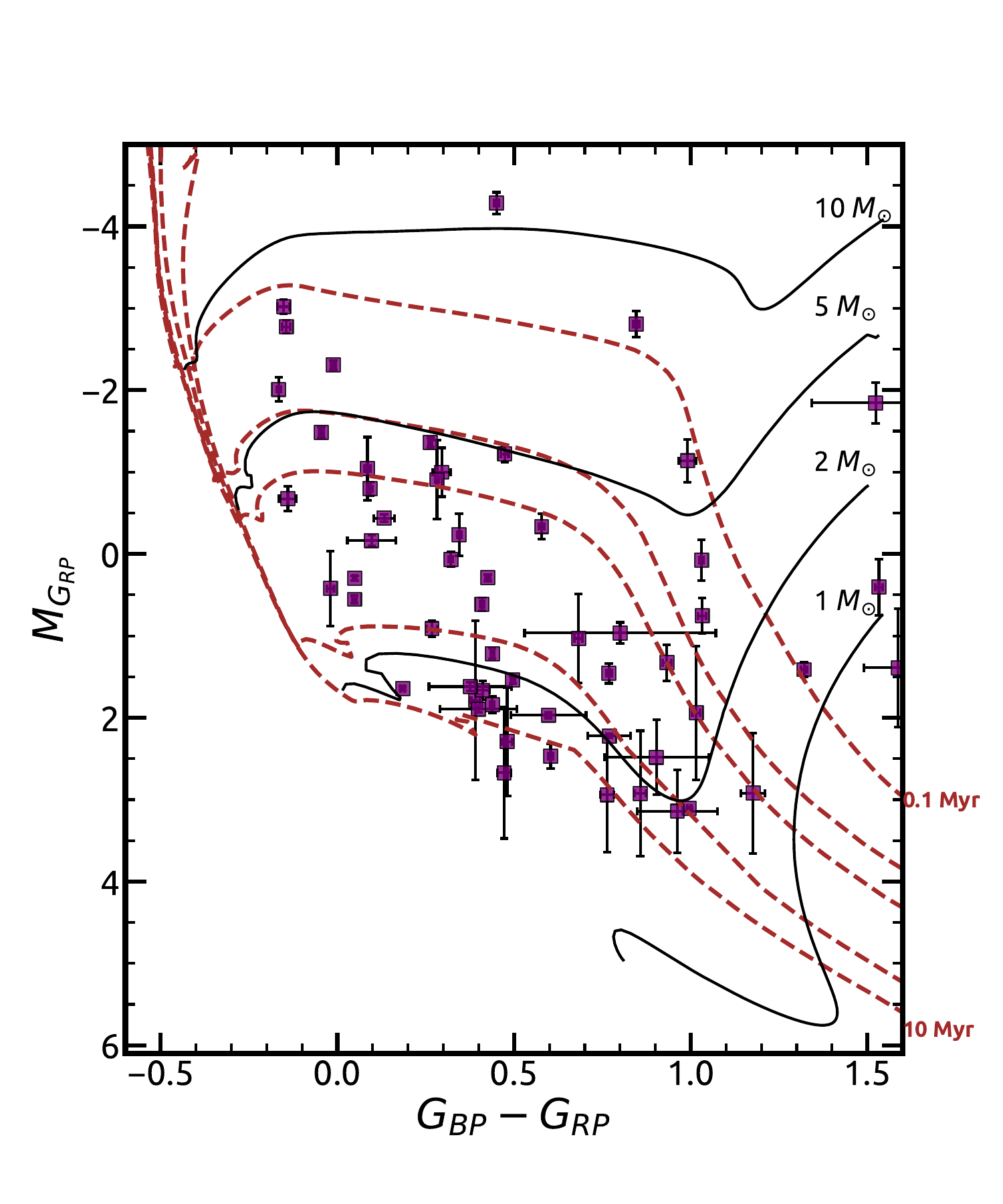}
    \caption{Figure represents the MIST isochrones and evolutionary tracks over \textit{Gaia} CMD for the HAeBe stars. The age for 61 HAeBe stars ranges from 0.1 till 10 Myr with $Z_{\odot}$ = 0.0512 and v/$v_{crit}$ = 0.4. The masses for 62 HAeBe stars range from 1.5 to 10 $M_{\odot}$. The isochrones of 0.1, 0.5, 1, 5, and 10 Myr are marked as brown dashed lines and the evolutionary tracks of 1, 2, 5, and 10 $M_{\odot}$ are marked as black solid lines.}
    \label{fig:age_mass}
\end{figure}

The age and mass of the stars are estimated by placing them in the \textit{Gaia} color-magnitude diagram (CMD) and over-plotted with the isochrones and theoretical evolutionary tracks, respectively. While constructing the \textit{Gaia} CMD, we used the broad-band photometric magnitudes, \textit{$G_{BP}$} and \textit{$G_{RP}$} from Gaia EDR3 \citep{2020yCat.1350....0G}. \textit{$G_{BP}$} and \textit{$G_{RP}$} are corrected for extinction (see section \ref{subsec:3.5}). Also, we made use of the Gaia distance estimates obtained from \cite{2021yCat.1352....0B} to calculate the absolute \textit{$G_{RP}$} magnitude ($M_{G_{RP}}$) using the distance modulus equation{\footnote{$m_{G_{RP_{0}}}$-$M_{G_{RP_{0}}}$ = 5log(d) - 5}}. Once the \textit{Gaia} CMD is constructed for the 119 HAeBe stars (see Figure \ref{fig:age_mass}), the age and mass are estimated using the updated models in the Modules of Experiments in Stellar Astrophysics (MESA) Isochrones and Stellar Tracks (MIST{\footnote{http://waps.cfa.harvard.edu/MIST/}}) \citep{2016ApJ...823..102C,2016ApJS..222....8D}. The updated models include the isochrones and evolutionary tracks for the metallicity $[Fe/H]$ = 0 (corresponding to solar metallicity; $Z_{\odot}$ = 0.0152) for Gaia EDR3, and v/$v_{crit}$ = 0.4.

We obtained the age for 61 stars from the Gaia CMD (Figure \ref{fig:age_mass}), among which five objects are lying above the 0.1 Myr isochrone. The estimated age of HAeBe stars ranges from 0.1 -– 10 Myr. Similarly, in Figure \ref{fig:age_mass}, the masses for 62 stars are estimated from the Gaia CMD, over-plotted with the evolutionary tracks. The estimated mass of the HAeBe stars range between 1.5 to 10 $M_{\odot}$. Most of the stars within the mass range of 2 $\leq$ $M_{\odot}$ $\leq$ 8 are younger than 3.5 Myr. In a few cases, the position of the stars is below the ZAMS and their age and mass are not estimated. Positions of these stars might be affected due to the photometric variability observed in HAeBe stars \citep{1998A&A...330..145V}.

\subsection{Estimation of Mass Accretion Rate}
\label{subsec:3.7}

\begin{figure*}
	\includegraphics[width=2.151\columnwidth]{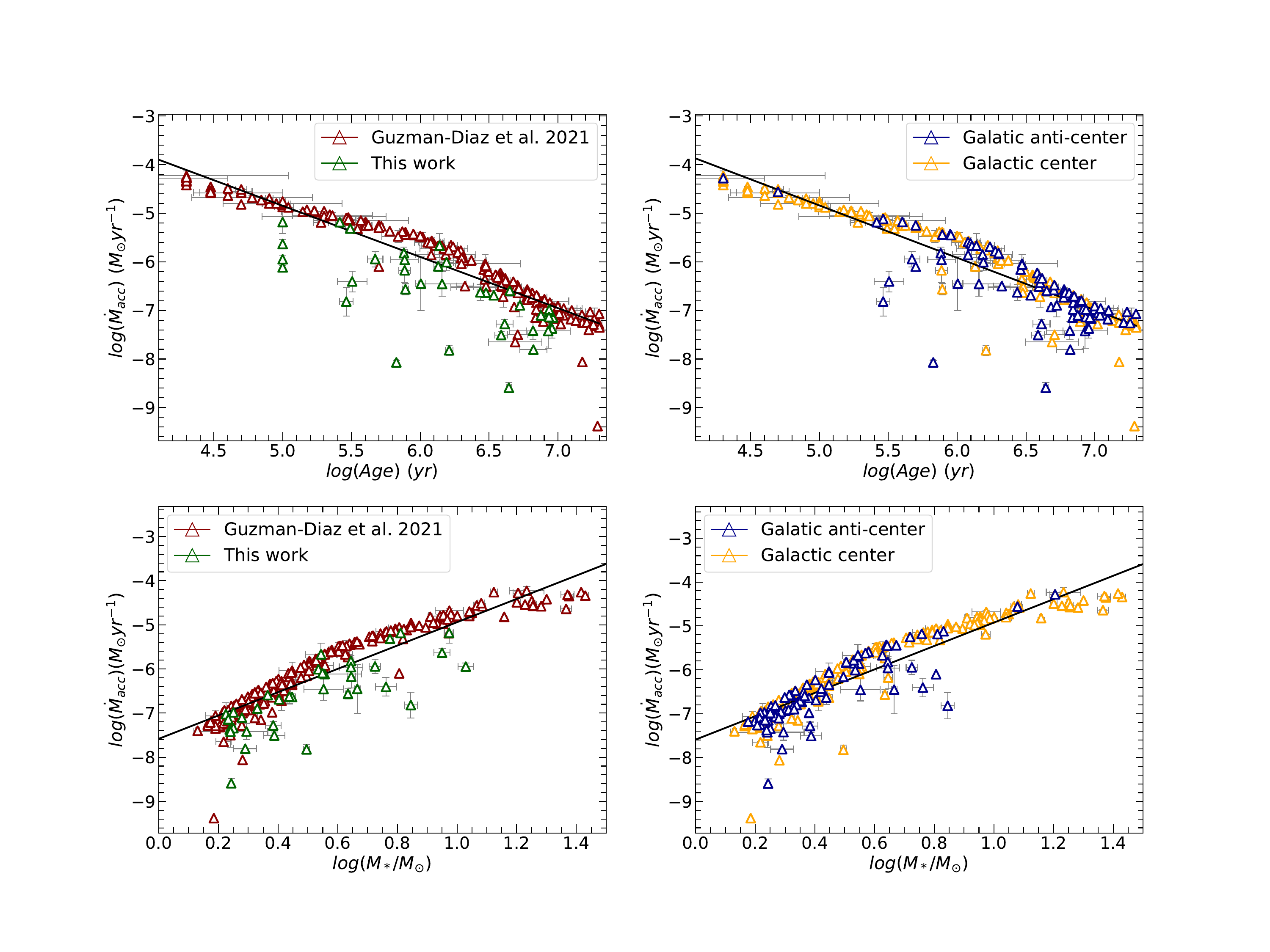}
    \caption{Figure illustrates the log-log plot of the derived $\dot{M}_{acc}$ against the age (top panel) and mass (bottom panel) of the star. Top left: The $\dot{M}_{acc}$ is found to be higher for younger stars. Bottom left: The $\dot{M}_{acc}$ is increasing with mass. Stars from this work are given in green and the sample from \protect\citet{2021A&A...650A.182G} is given in brown color. Top right and Bottom right: Similar trend of higher $\dot{M}_{acc}$ for younger and massive stars are found in the Galactic center (orange) and anti-center directions (blue). The solid black line represents the corresponding best fit to the data. The slope of the best fits for the whole sample in the top (both right and left) panel are -1.05 $\pm$ 0.05 and -1.08 $\pm$ 0.17, respectively. Similarly, the slope of the best fits for the whole sample in the bottom (both right and left) panel are 2.74 $\pm$ 0.54 and 2.77 $\pm$ 0.21, respectively.  } 
    \label{fig:Macc}
\end{figure*}

The distinct mass accretion modes in HAeBe stars is an ongoing study in the research community. Many evidence point out towards the boundary layer (BL) accretion mode for early-HBe stars while magnetospheric accretion (MA) is suitable for HAe stars \citep{2020MNRAS.493..234W, 2020Galax...8...39M}. However, due to low number of early B-type stars in our sample, and no valid method is devised to calculate $\dot{M}_{acc}$ using the BL mode, we are estimating the $\dot{M}_{acc}$ using the MA technique. In this work, we estimate the mass accretion rate of HAeBe stars and analyze the possible range of mass accretion rates and its significance.

An empirical correlation is derived between the accretion luminosities and the line luminosities, which also correlate with $\dot{M}_{acc}$ for PMS stars \citep{2011A&A...529A..34M, 2015MNRAS.452.2837M, 2017MNRAS.464.4721F}. Thus, in this work, we measure the EW of the strongest line present in the spectra, ie., the EW(H$\alpha$), to estimate the accretion luminosity. The H$\alpha$ line flux $(F_{H\mathrm{\alpha}})$ is estimated by $(F_{H\mathrm{\alpha}}) = W(H\mathrm{\alpha}) \times  F_{R}(H\mathrm{\alpha})$, where $F_{R}(H\mathrm{\alpha})$ is the continuum flux density at H$\alpha$. The $F_{R}(H\mathrm{\alpha})$ is calculated using the extinction corrected \textit{R} band magnitude ($R_0$, \citealp{2018ApJ...857...30M}). Expanding upon this, the $F_{H\mathrm{\alpha}}$ is converted to line luminosity using the relationship, $L_{H\mathrm{\alpha}}=4\pi D^2 F_{H\mathrm{\alpha}}$, where D is the distance to the star compiled from \cite{2021yCat.1352....0B}. We used the empirical relationship between line luminosity and accretion luminosity ($L_{acc}$) used by \cite{2017MNRAS.464.4721F} and \cite{2019yCat..51570159A}, to calculate the $L_{acc}$ of the HAeBe stars in our sample. The relationship is given by, 

\begin{equation}
   \log\left(\frac{L_{\mathrm{acc}}}{L_{\odot}}\right) = 2.09 (\pm 0.06) + 1.00 (\pm 0.05) \times \log\left(\frac{L_{H\alpha}}{L_{\odot}}\right)
\end{equation}

The $\dot{M}_{acc}$ is then calculated from the following equation.
\begin{equation}
L_{acc} = \frac{GM_*\dot{M}_{acc}}{R_*}(1-\frac{R_*}{R_i})
\end{equation}

where $M_*$ is the mass of HAeBe stars, $R_*$ is the stellar radius, and $R_i$ is the disc truncation radius. For HAeBe stars, the disc truncation radius is fixed to be at 2.5 $R_*$, which should be lower than the co-rotation radius ($R_{cor}$) where the matter is accreted towards the star through the magnetic field lines \citep{1994ApJ...429..781S, 2004ApJ...617..406M, 2011A&A...535A..99M}. $\dot{M}_{acc}$ for 40 HAeBe stars are estimated and are found to be in the range $10^{-6}$ -- $10^{-9}$ $M_{\odot}yr^{-1}$.

\subsubsection{Mass accretion rate as function of stellar age and mass}
\label{subsubsec:3.7.1}

 The relationship of mass accretion rate with the age and mass of HAeBe stars is shown in Figure \ref{fig:Macc}. Massive stars have a lower pre-main sequence lifetime than low-mass stars and hence will be younger. In the case of HAeBe stars, majority of the HAe stars will be older than HBe stars. The top left panel of Figure \ref{fig:Macc} shows a decrease in accretion rate with an increase in the age of the HAeBe stars. Based on the relationship of $\dot{M}_{acc}$ $\propto$ $t^{-\eta}$, where t is the age in Myr, we obtained $\eta$ value as 1.1 $\pm$ 0.2. This is in agreement with the literature as shown in Figure \ref{fig:Macc}, where \cite{2021A&A...650A.182G} found $\eta$ = 1.03 $\pm$ 0.02. Similarly, \cite{2012A&A...543A..59M}, \cite{2015MNRAS.453..976F}, and \cite{2019yCat..51570159A} estimated $\eta$ values to be ${1.8^{+1.4}_{-0.7}}$, 1.92 $\pm$ 0.09, and 1.2 $\pm$ 0.1 respectively. We can see that $\eta$ values estimated by \cite{2012A&A...543A..59M} and \cite{2015MNRAS.453..976F} is relatively higher than that determined in this study. 

The accretion rates are largest for the brightest and most massive objects. Similarly, the correlation between log($\dot{M}_{acc}$) and stellar mass for HAeBe stars is plotted in the bottom left panel of Figure \ref{fig:Macc}. We obtained $\dot{M}_{acc}$ $\propto$ $M_{*}^{3.12^{+0.21}_{-0.34}}$ as the best fit to the data from this work, which is agreement to $\dot{M}_{acc}$ $\propto$ $M_{*}^{2.8\pm0.2}$ obtained by \cite{2019yCat..51570159A}. In Figure \ref{fig:Macc}, we compared the sample with the \cite{2021A&A...650A.182G} sample, where the stars are found to follow the similar trend. Also in the literature, a steeper power-law of $\dot{M}_{acc}$ $\propto$ $M_{*}^{5}$ by \cite{2011A&A...535A..99M} and  $\dot{M}_{acc}$ $\propto$ $M_{*}^{3.72}$ by \cite{2015MNRAS.453..976F} is obtained. The difference in the relations is primarily due to the difference in sample size and sample mass ranges. Hence, we do not find much difference in the power-law index relating the mass accretion rate with the mass of HAeBe stars. Also, the top and bottom right panels of Figure \ref{fig:Macc} show that the dependency of $\dot{M}_{acc}$ with age and mass is valid for HAeBe stars lying in both the Galactic center and anti-center directions.

\section{Conclusion}
\label{sec:4}
    
In this work, we carefully studied a sample of 119 HAeBe stars
identified from LAMOST DR5. The initial sample of 3339 ELS stars was obtained from \citetalias{2021RAA....21..288S}, where they classified a sample of OBA ELS in the Galaxy into different categories based on the evolutionary phase. Further, a sample of F-type ELS was included (after following the similar classification method outlined in \citetalias{2021RAA....21..288S}), since HAeBe stars span a wide spectral range up to F5, making the total number of ELS to 3850. The selection criteria included the presence of H$\alpha$ emission, near-IR color values (\textit{H}-\textit{K$_s$})$_0$ > 0.4 and infrared spectral index \textit{$n_{(K_{s}-W2)}$} > -1.5, resulting in a sample of 151 possible HAeBe stars. A thorough visual check of 151 HAeBe stars showed that 32 stars exhibit nebular forbidden emission lines of [O{\sc iii}], which are removed. Finally, the sample of HAeBe stars used for this study is 119. It is found that out of 119 stars, 51 stars were confirmed HAeBe, 12 as candidate HAeBe and 30 as H$\alpha $ ELS by \cite{2022ApJS..259...38Z}. Additionally, 5 F-type stars are identified as HAeBe by \cite{2022ApJ...936..151Z}. Also, according to the SIMBAD database, five stars are studied as YSO/star in well-known studies of YSOs, and two stars are labeled as Mira variable and evolved star, respectively. However, from the age and mass estimates and the location in the CMD, we found that they are HAeBe stars. Additionally, they do not show any nebular lines or high-ionization emission lines observed in Mira variables or evolved stars. Therefore, these two stars are also reclassified as HAeBe stars and are included in our sample. The remaining 14 stars are not yet studied in the literature. The HAeBe nature of this new sample of stars and those 30 ELS reported by \cite{2022ApJS..259...38Z} is confirmed through our study. The other important results from this study are summarised below.

\begin{itemize}

    \item Due to the observational strategy of LAMOST, the HAeBe stars identified in this study are found to be distributed in the Galactic plane towards the Galactic anti-center direction. This is quite different from the location of all other known HAeBe stars from the literature. No considerable amount of information is available in the literature regarding the HAeBe stars spread across the Galactic anti-center direction. We used a template matching method using MILES stellar library to estimate the spectral type of HAeBe stars \citep{2006MNRAS.371..703S}.
     
    \item The visual check performed on all the 119 HAeBe stellar spectra helped to identify the major emission lines including H$\alpha$, Fe{\sc ii}, Ca{\sc II} triplet and forbidden lines. Apart from emission lines, several absorption features are observed, such as He{\sc i}, Na{\sc i} doublet, Mg{\sc ii}, DIB, and other metallic lines. According to the nature of the H$\alpha$ line profiles, the spectra are classified into three distinct classes. 
    
    \item The equivalent width of spectral lines such as H$\alpha$, Fe{\sc ii}, He{\sc i}, DIB, Ca{\sc ii} triplet, and forbidden lines including [O{\sc i}] 6300 \AA, 6363 \AA, [N{\sc ii}] 6548 \AA, 6584 \AA~and [S{\sc ii}] 6717 \AA, 6731 \AA~are measured and tabulated in Table \ref{tab:A1}. It is observed that the EW(H$\alpha$) is higher for early-type Herbig Be stars, which subsequently decreased for late-type Herbig Ae stars. The He{\sc i} 5876 \AA~line ceases to appear in the stars later than A3 type.
    
    \item  The Fe{\sc ii} 5169 \AA~emission region could be in the inner disc, with an overlap with the H$\alpha$ emitting region, due to which we find a moderate correlation between the emission strengths. Also, we do not see any correlation between Fe{\sc ii} 7712 \AA~and H$\alpha$. Hence, it is evident from the present analysis that, Fe{\sc ii} emission line is due to the contribution from various emission regions around the star. Further studies in the future will be required to identify the mechanism and the exact region of formation of Fe{\sc ii} lines in the HAeBe stars.
    
    \item  Information regarding the extinction values for majority of the stars is not readily available in the literature. As an alternative, we employed a new method using the interstellar spectral feature known as DIB to correct for extinction in a sample of HAeBe stars. Equivalent width of DIB present in 93 HAeBe stars are measured and converted into reddening using the relation given by \cite{2011ApJ...727...33F}. Based on the calculated reddening values, we estimated the extinction for 93 HAeBe stars. For rest of the stars without DIB features, the extinction values are estimated using the extinction map of \cite{2019ApJ...887...93G}.
    
    \item Using \textit{Gaia} CMD, we determined the age and mass of HAeBe stars ranging between 0.1 -- 10 Myr and 1.5 -- 10 $M_{\odot}$, respectively. Further, the mass accretion rates are derived which remain within a range of $10^{-6}$ -- $10^{-9}$ $M_{\odot}yr^{-1}$. We find a clear trend of mass accretion rates increasing as a function of stellar mass. Irrespective of the location of the stars, either in Galactic center or Galactic anti-center direction, the power-law relationship obtained for our stars ($\dot{M}_{acc}$ $\propto$ $M_{*}^{3.12^{+0.21}_{-0.34}}$) is found to be in agreement with the literature \citep{2011A&A...535A..99M, 2015MNRAS.453..976F, 2018A&A...620A.128V, 2019yCat..51570159A}. The relation between mass accretion rates with the age of our stars, given by $\dot{M}_{acc}$ $\propto$ $t^{-\eta}$, is obtained in the form of $\dot{M}_{acc} \propto t^{-1.1 \pm 0.2}$, confirming that younger stars have higher mass accretion rates. Also, the dependency of mass accretion rate with age and mass is found to be valid for HAeBe stars lying towards both the Galactic center and anti-center directions.

\end{itemize}

\section*{Acknowledgements}

We would like to thank our referees for providing helpful comments and suggestions that improved the paper. We thank Robin Thomas for his valuable suggestion throughout the course of the work. This work was financially supported by the Management of CHRIST (Deemed to be University), Bangalore. The authors are grateful to the Centre for Research, CHRIST (Deemed to be University), Bangalore for the research grant extended to carry out the project (MRPDSC-1932). Also, we would like to thank the Science \& Engineering Research Board (SERB), a statutory body of the Department of Science \& Technology (DST), Government of India, for funding our research under grant number CRG/2019/005380. RA acknowledges the financial support from SERB POWER fellowship grant SPF/2020/000009. This work has made use of data products from the Guo Shoujing Telescope (the Large Sky Area Multi-Object Fibre Spectroscopic Telescope, LAMOST), and data from the European Space Agency (ESA) mission Gaia (https://www.cosmos.esa.int/gaia), processed by the Gaia Data Processing and Analysis Consortium (DPAC, https://www.cosmos.esa.int/web/gaia/dpac/consortium). Funding for the DPAC has been provided by national institutions, in particular the institutions participating in the Gaia Multilateral Agreement. We thank the SIMBAD database and the online VizieR library service for helping us in the literature survey and obtaining relevant data.
\section*{Data Availability}

The data underlying this article were accessed from the Large sky Area Multi-Object fibre Spectroscopic Telescope (LAMOST) data release 5 (http://dr5.lamost.org/). The derived data generated in this research will be shared on a reasonable request to the corresponding author.

\bibliographystyle{mnras}
\bibliography{Herbig} 
\onecolumn
\begin{landscape}
\tiny
\begin{longtable}[c]{lllllllllllllllll}
\caption{The measured equivalent widths (EW) of H$\alpha$, Ca{\sc ii} triplet, and the four DIB used to calculate the reddening of the HAeBe stars is listed in columns 2 to 9. The measured EW of Fe{\sc ii} 5169 \AA, He{\sc i} 5876 \AA, and forbidden lines including [O{\sc i}] 6300 \AA, 6363 \AA, [N{\sc ii}] 6548 \AA, 6584 \AA~and [S{\sc ii}] 6717 \AA, 6731 \AA~are listed in columns 10 to 17. Equivalent widths of H$\alpha$ and Fe{\sc ii} 5169 \AA~are photospherically absorption corrected. (-) sign in all the columns denote emission whereas positive values denote absorption. Upper limits of the equivalent widths for those lines that were undetected or appeared as minor peaks are indicated with an asterisk ($*$).}
\label{tab:A1}\\
\hline\hline
 LAMOST\_ID  & \multicolumn{16}{c}{Equivalent Width (\AA)} \\

 & H$\alpha$ & Ca{\sc ii} & Ca{\sc ii} & Ca{\sc ii} & DIB & DIB & DIB & DIB & Fe{\sc ii} & He{\sc i} & [O{\sc i}] & [O{\sc i}] & [N{\sc ii}] & [N{\sc ii}] & [S{\sc ii}] & [S{\sc ii}]              \\
 & 6563 \AA & 8498 \AA & 8542 \AA & 8662 \AA & 5780 \AA & 5797 \AA & 6614 \AA & 6284 \AA & 5169 \AA & 5876 \AA & 6300 \AA & 6363 \AA & 6548 \AA & 6584 \AA & 6717 \AA & 6731 \AA \\
\hline\hline
\endfirsthead

\hline\hline

LAMOST\_ID  & \multicolumn{16}{c}{Equivalent Width (\AA)} \\

 & H$\alpha$ & Ca{\sc ii} & Ca{\sc ii} & Ca{\sc ii} & DIB & DIB & DIB & DIB & Fe{\sc ii} & He{\sc i} & [O{\sc i}] & [O{\sc i}] & [N{\sc ii}] & [N{\sc ii}] & [S{\sc ii}] & [S{\sc ii}]              \\
 & 6563 \AA & 8498 \AA & 8542 \AA & 8662 \AA & 5780 \AA & 5797 \AA & 6614 \AA & 6284 \AA & 5169 \AA & 5876 \AA & 6300 \AA & 6363 \AA & 6548 \AA & 6584 \AA & 6717 \AA & 6731 \AA \\
\hline\hline
\endhead

J025146.96+554201.3 & -24.8$\pm$1.1 & -4.6$\pm$0.1 & -4.3$\pm$0.1 & -4.1$\pm$$\le$0.1 & 0.6$\pm$0.2 &  &  & 1.4$\pm$0.1 & -1.0$\pm$0.1 &  &  &  &  &  &  &  \\
J032832.62+511354.4 & -10.1$\pm$0.5 & -0.9$\pm$0.1* & -1.4$\pm$0.1* & -1.0$\pm$$\le$0.1* & 0.5$\pm$$\le$0.1 & 0.2$\pm$$\le$0.1 &  &  &  & $\le$-0.1$\pm$$\le$0.1* &  &  &  &  &  &  \\
J032907.54+570133.7 &  & -3.0$\pm$0.1 & -4.5$\pm$0.1 & -3.6$\pm$$\le$0.1 & 1.2$\pm$0.2 & 0.3$\pm$$\le$0.1 &  &  &  &  & -6.2$\pm$0.1 & -2.3$\pm$0.1 &  &  &  &  \\
J033900.56+294145.7 & -25.4$\pm$1.2 &  &  &  & 0.1$\pm$$\le$0.1 &  &  & 1.1$\pm$0.1 & -0.2$\pm$$\le$0.1* &  &  &  &  &  &  &  \\
J034109.13+314437.8 &  & 0.8$\pm$0.2 & 1.9$\pm$0.1 & 1.7$\pm$0.3 & 0.5$\pm$0.1* & 0.8$\pm$0.1* &  & 0.4$\pm$$\le$0.1 &  &  &  &  &  &  &  &  \\
J034255.96+315841.7 & -23.2$\pm$1.6 & 0.9$\pm$0.1 & 1.4$\pm$$\le$0.1 & 3.2$\pm$0.9 & 0.6$\pm$0.1* &  &  & 0.4$\pm$$\le$0.1 &  & 1.0$\pm$0.1* &  &  &  &  &  &  \\
J035553.38+363840.4 & -60.5$\pm$1.4 &  &  &  &  &  &  & 0.7$\pm$0.1 &  &  &  &  &  & -3.2$\pm$0.1 & -2.6$\pm$$\le$0.1 & -1.6$\pm$$\le$0.1 \\
J035952.31+481339.8 & -24.8$\pm$$\le$0.1 & -2.5$\pm$$\le$0.1 & -2.6$\pm$0.4 & -2.6$\pm$0.3 & 0.7$\pm$0.1 & 0.2$\pm$$\le$0.1 & 0.2$\pm$$\le$0.1* & 1.9$\pm$0.1 & -1.5$\pm$$\le$0.1 & 0.5$\pm$$\le$0.1 &  &  &  &  &  &  \\
J040058.11+572247.7 & -33.9$\pm$0.1 & -4.6$\pm$0.2 & -5.0$\pm$$\le$0.1 & -3.5$\pm$0.2 & 0.2$\pm$0.1 &  & 0.2$\pm$$\le$0.1 & 1.4$\pm$$\le$0.1 & -0.7$\pm$$\le$0.1 & 0.6$\pm$0.1 &  &  &  &  &  &  \\
J040559.62+295638.1 & -9.0$\pm$0.7 & 1.0$\pm$0.1 & 1.1$\pm$0.2 & 1.6$\pm$1.8 & $\le$0.1$\pm$$\le$0.1* & $\le$0.1$\pm$$\le$0.1* &  & 1.2$\pm$0.2 & -0.1$\pm$$\le$0.1 &  & -0.3$\pm$0.1 &  &  &  &  &  \\
J041150.68+503619.6 & -36.5$\pm$0.3 & -1.8$\pm$0.1 & -1.9$\pm$0.1 & -1.7$\pm$0.1 & 0.6$\pm$0.1 & 0.2$\pm$$\le$0.1 &  & 1.0$\pm$0.1 & -1.4$\pm$$\le$0.1* & 0.5$\pm$$\le$0.1 &  &  &  & -0.5$\pm$$\le$0.1* &  &  \\
J041805.77+533707.2 & -21.6$\pm$1.4 & -1.7$\pm$0.4 & -1.5$\pm$$\le$0.1 & -1.1$\pm$$\le$0.1 & 0.6$\pm$$\le$0.1 & 0.2$\pm$0.1 &  & 1.2$\pm$0.1 & -0.2$\pm$$\le$0.1 & 0.1$\pm$$\le$0.1* &  &  &  &  &  &  \\
J042622.24+440031.4 & -55.5$\pm$2.4 & -4.4$\pm$0.2 & -4.8$\pm$$\le$0.1 & -4.4$\pm$$\le$0.1 & 0.3$\pm$0.1 &  & 0.2$\pm$$\le$0.1* & 1.6$\pm$0.1 & -1.4$\pm$$\le$0.1 & 0.6$\pm$0.1 &  &  &  &  &  &  \\
J042853.11+483132.4 & -34.1$\pm$1.1 &  &  &  & 0.4$\pm$0.1 & 0.1$\pm$$\le$0.1 &  & 0.9$\pm$0.1 & -0.1$\pm$$\le$0.1 & 0.7$\pm$0.2 &  &  &  &  &  &  \\
J043000.64+422259.6 & -58.3$\pm$1.9 & -5.6$\pm$$\le$0.1 & -5.4$\pm$$\le$0.1 & -4.2$\pm$$\le$0.1 & 0.2$\pm$$\le$0.1 &  & 0.5$\pm$0.1* & 0.8$\pm$0.1 & -1.4$\pm$0.1 & 0.5$\pm$0.1 &  &  &  &  &  &  \\
J043718.84+450641.9 & -10.8$\pm$0.5 & 0.4$\pm$$\le$0.1 & 1.0$\pm$0.1 & 1.2$\pm$0.1 & 0.4$\pm$$\le$0.1 &  &  & 0.7$\pm$0.1 & -0.2$\pm$0.1 &  &  &  &  &  &  &  \\
J050303.79+422144.6 & -39.6$\pm$3.0 &  &  &  & 0.4$\pm$0.1 &  &  & 1.3$\pm$0.2 &  &  &  &  &  &  &  &  \\
J050924.57+520007.1 & -26.5$\pm$0.2 &  &  &  & 0.4$\pm$0.1 &  &  & 1.6$\pm$$\le$0.1 & -0.1$\pm$0.2 & 0.5$\pm$0.1* &  &  &  &  &  &  \\
J051039.20+305419.6 & -14.1$\pm$0.9 &  &  &  & 0.3$\pm$0.1 &  &  & 1.3$\pm$0.1 & -0.2$\pm$0.1 & 0.6$\pm$$\le$0.1 &  &  &  &  &  &  \\
J051133.52+334849.7 &  &  &  &  & 0.4$\pm$$\le$0.1 &  &  & 0.5$\pm$0.1 &  &  &  &  &  &  &  &  \\
J051159.17+342843.0 &  &  &  &  &  &  &  &  &  &  & -1.2$\pm$0.1* &  & -4.4$\pm$$\le$0.1 & -14.9$\pm$1.4 & -6.3$\pm$$\le$0.1 & -4.2$\pm$$\le$0.1 \\
J051425.20+411310.7 & -89.2$\pm$2.8 &  &  &  & 0.1$\pm$$\le$0.1* & 0.1$\pm$$\le$0.1* &  &  & 0.0$\pm$$\le$0.1* & 0.8$\pm$0.1* &  &  &  &  &  &  \\
J051439.95+324030.5 & -32.5$\pm$0.8 &  &  &  & 0.6$\pm$0.4* & 0.3$\pm$0.1* &  & 1.7$\pm$$\le$0.1 & -0.1$\pm$$\le$0.1* &  &  &  &  &  &  &  \\
J051634.39+402856.1 &  &  &  &  & 0.5$\pm$0.1* &  &  & 0.9$\pm$0.1 &  &  & -0.8$\pm$0.1* & -0.4$\pm$$\le$0.1* &  &  &  &  \\
J051727.81+344343.0 &  &  &  &  &  & 1.0$\pm$0.1* &  & 1.4$\pm$0.4 &  &  & -0.7$\pm$0.1 &  & -0.6$\pm$0.1* & -1.9$\pm$0.4* & -1.7$\pm$3.0 & -0.5$\pm$0.1 \\
J052017.50+391657.8 & -14.0$\pm$0.7 & -2.6$\pm$0.1 & -4.3$\pm$0.1 & -3.5$\pm$$\le$0.1 & 0.5$\pm$$\le$0.1 & 0.1$\pm$$\le$0.1 &  & 1.3$\pm$0.1 & -0.5$\pm$0.1 & 0.1$\pm$$\le$0.1* &  &  &  &  &  &  \\
J052019.20+320817.0 & -30.1$\pm$1.3 & -3.8$\pm$0.2 & -3.9$\pm$0.2 & -3.6$\pm$0.1 & 0.7$\pm$$\le$0.1 & 0.2$\pm$0.1 & 0.2$\pm$$\le$0.1* & 1.8$\pm$0.1 & -1.5$\pm$$\le$0.1 & 0.2$\pm$$\le$0.1 &  &  &  &  &  &  \\
J052033.42+330857.6 & -37.0$\pm$1.4 &  &  &  & 0.6$\pm$0.1* & 1.5$\pm$0.2* &  &  &  &  & -3.2$\pm$0.4* &  &  &  &  &  \\
J052050.38+315637.5 & -35.4$\pm$0.8 &  &  &  &  &  &  & 0.7$\pm$0.1 & -0.9$\pm$$\le$0.1* &  &  &  &  & -0.5$\pm$0.1 &  &  \\
J052058.21+041827.6 & -11.6$\pm$0.1 &  &  &  &  &  &  & 1.2$\pm$0.2 &  &  &  &  &  &  &  &  \\
J052135.30-045329.2 & -18.1$\pm$0.7 & -2.4$\pm$$\le$0.1 & -1.5$\pm$$\le$0.1 & -1.4$\pm$0.2 &  &  & 0.5$\pm$0.1 & 2.2$\pm$0.6 &  & -0.3$\pm$$\le$0.1 &  &  &  &  &  &  \\
J052250.18+413711.9 & -13.5$\pm$0.4 &  &  &  &  &  &  & 1.6$\pm$$\le$0.1 &  & -0.8$\pm$-0.1* &  &  &  &  &  &  \\
J052322.02+281011.3 &  &  &  &  & 0.9$\pm$0.1 &  &  &  &  &  & -4.1$\pm$0.1 & -1.8$\pm$0.1 &  &  &  &  \\
J052330.87+051726.8 & -15.4$\pm$0.5 &  &  &  &  &  &  &  &  &  &  &  &  &  &  &  \\
J052707.42+341822.7 & -17.7$\pm$0.7 &  &  &  &  &  &  & 1.1$\pm$0.1 & -0.2$\pm$$\le$0.1* &  &  &  &  &  &  &  \\
J052919.14+341747.1 & -65.0$\pm$1.3 & -1.9$\pm$0.1 & -2.4$\pm$0.2 & -2.5$\pm$0.1 & 0.4$\pm$$\le$0.1 &  & 0.1$\pm$$\le$0.1* & 0.9$\pm$0.2 & 0.0$\pm$$\le$0.1* & 0.7$\pm$0.2 &  &  &  &  &  &  \\
J052926.22+351610.5 & -36.4$\pm$0.8 &  &  &  & 0.7$\pm$$\le$0.1 &  &  & 0.6$\pm$$\le$0.1 &  &  &  &  &  &  &  &  \\
J052947.46+332915.5 & -27.2$\pm$0.2 &  &  &  & 0.4$\pm$0.1 &  &  & 1.5$\pm$0.2 & -0.1$\pm$$\le$0.1* & 0.4$\pm$0.1* &  &  &  &  &  &  \\
J052948.05-002343.4 & -13.9$\pm$0.4 & 0.9$\pm$$\le$0.1 & 1.3$\pm$0.3 & 3.5$\pm$$\le$0.1 &  &  &  & 1.5$\pm$0.1 & -0.1$\pm$$\le$0.1 &  & -0.8$\pm$0.1 &  &  & -1.2$\pm$1.4 & -1.4$\pm$1.3 &  \\
J053024.01+373010.2 &  & -3.3$\pm$0.3 & -3.6$\pm$$\le$0.1 & -3.5$\pm$0.2 & 0.4$\pm$$\le$0.1 &  &  &  &  &  &  &  &  &  &  &  \\
J053123.43+383157.1 & -69.5$\pm$2.1 & -3.1$\pm$$\le$0.1 & -2.9$\pm$$\le$0.1 & -2.8$\pm$$\le$0.1 & 0.8$\pm$0.1 &  &  &  &  &  &  &  &  &  &  &  \\
J053148.50+152211.7 & -11.3$\pm$0.8 &  &  &  &  &  & 0.2$\pm$$\le$0.1 & $\le$0.1$\pm$3.6 &  &  &  &  &  &  &  &  \\
J053200.30-045553.8 &  &  &  &  &  &  &  & 0.3$\pm$0.1 &  &  & $\le$0.1$\pm$0.6 &  &  &  &  &  \\
J053209.94-024946.7 & -8.4$\pm$0.2 & 1.0$\pm$$\le$0.1 & 2.5$\pm$$\le$0.1 & 2.1$\pm$$\le$0.1 &  &  &  & 1.1$\pm$0.1 & -0.2$\pm$$\le$0.1 &  &  &  &  &  &  &  \\
J053637.36+330003.9 & -7.4$\pm$1.3 & 1.0$\pm$$\le$0.1 & 1.3$\pm$$\le$0.1 & 1.5$\pm$0.3 & 0.3$\pm$$\le$0.1 & 0.6$\pm$0.7 &  & 0.7$\pm$0.1 & -0.2$\pm$0.1 &  &  &  &  &  &  &  \\
J053857.22+373003.2 & -30.8$\pm$0.3 &  &  &  & 0.4$\pm$$\le$0.1 & 0.2$\pm$$\le$0.1 &  & 0.6$\pm$0.1 &  & -0.6$\pm$$\le$0.1* & -0.5$\pm$$\le$0.1 &  &  &  &  &  \\
J053918.09+361716.2 & -25.9$\pm$0.2 & -8.7$\pm$0.1 & -5.8$\pm$0.3 & -5.0$\pm$$\le$0.1 & 0.5$\pm$$\le$0.1 & 0.2$\pm$$\le$0.1 & 0.2$\pm$$\le$0.1 & 1.7$\pm$$\le$0.1 & -1.1$\pm$$\le$0.1 & 0.4$\pm$0.1* &  &  &  &  &  &  \\
J054109.42+360836.8 & -36.4$\pm$1.0 &  &  &  & 0.7$\pm$0.1* &  & 0.3$\pm$$\le$0.1* & 1.2$\pm$$\le$0.1 & -0.4$\pm$$\le$0.1* &  & -0.4$\pm$$\le$0.1 &  &  &  &  &  \\
J054113.78+273938.5 &  &  &  &  &  &  &  & 1.5$\pm$0.1 &  &  &  &  &  &  &  &  \\
J054252.77+114401.9 &  &  &  &  &  &  &  &  &  &  &  &  & -1.5$\pm$$\le$0.1 & -2.1$\pm$$\le$0.1 & -1.3$\pm$0.1 & -1.0$\pm$$\le$0.1 \\
J054329.26+000458.8 & -15.4$\pm$0.2 & 1.8$\pm$0.3 & 3.8$\pm$0.7 & 3.9$\pm$$\le$0.1 &  &  &  & 0.6$\pm$$\le$0.1 & -1.6$\pm$0.2 &  &  &  &  &  &  &  \\
J054424.58+342802.7 & -23.0$\pm$1.0 & -4.7$\pm$0.3 & -2.7$\pm$0.1 & -1.9$\pm$0.3 & 0.6$\pm$$\le$0.1 & 0.2$\pm$$\le$0.1 &  & 0.5$\pm$0.2 & -0.3$\pm$$\le$0.1 &  &  &  &  &  &  &  \\
J054431.95+320449.2 & -14.3$\pm$0.2 & -0.5$\pm$0.1* & -0.8$\pm$$\le$0.1* & -0.9$\pm$0.1* & 0.3$\pm$$\le$0.1 &  &  & 1.2$\pm$$\le$0.1 &  &  &  &  &  &  &  &  \\
J054454.86+215736.4 & -9.7$\pm$0.5 & 0.9$\pm$$\le$0.1 & 2.0$\pm$$\le$0.1 & 1.5$\pm$$\le$0.1 & 0.3$\pm$$\le$0.1 &  & 0.2$\pm$$\le$0.1* & 1.1$\pm$$\le$0.1 & -0.3$\pm$0.1 &  &  &  &  &  &  &  \\
J054649.95+220630.2 &  &  &  &  &  &  &  & 1.0$\pm$$\le$0.1 &  &  &  &  &  &  &  &  \\
J054658.35+304125.4 & -28.7$\pm$0.4 & -1.2$\pm$$\le$0.1 & -1.7$\pm$$\le$0.1 & -0.9$\pm$0.1 & 0.4$\pm$$\le$0.1 & 0.4$\pm$0.1 &  & 1.8$\pm$0.3 & 0.0$\pm$$\le$0.1* & 0.2$\pm$$\le$0.1* & -0.2$\pm$$\le$0.1* &  &  &  &  &  \\
J055054.77+201447.6 &  & -8.9$\pm$0.1 & -9.7$\pm$0.1 & -4.4$\pm$5.9 & 0.3$\pm$0.1 &  &  &  &  & 0.6$\pm$$\le$0.1 &  &  &  &  &  &  \\
J055530.50+115420.6 & -13.7$\pm$0.6 &  &  &  &  &  &  & 1.0$\pm$0.3 & -0.1$\pm$$\le$0.1 &  &  &  &  &  &  &  \\
J055547.52+174350.0 & -13.8$\pm$1.0 & 1.6$\pm$$\le$0.1 & 2.3$\pm$0.2 & 3.1$\pm$$\le$0.1 & 0.3$\pm$$\le$0.1 & 0.1$\pm$$\le$0.1 &  & 0.7$\pm$$\le$0.1 & -0.2$\pm$0.1 &  &  &  &  &  &  &  \\
J055831.50+264236.0 & -39.6$\pm$1.0 & -3.9$\pm$0.3 & -4.0$\pm$$\le$0.1 & -3.3$\pm$0.1 & 0.5$\pm$$\le$0.1 &  & 0.2$\pm$$\le$0.1* & 0.9$\pm$0.1 & -0.5$\pm$$\le$0.1* & 0.7$\pm$0.1* &  &  &  &  &  &  \\
J055838.98+201108.5 & -7.0$\pm$0.5 & 0.5$\pm$$\le$0.1 & 0.7$\pm$$\le$0.1 & 3.1$\pm$0.7 & 0.3$\pm$$\le$0.1 & 0.1$\pm$$\le$0.1 &  & 1.1$\pm$0.1 & -0.1$\pm$$\le$0.1 & 0.6$\pm$$\le$0.1 &  &  &  &  &  &  \\
J055910.90+120215.0 &  &  &  &  & 0.5$\pm$$\le$0.1 &  &  & 1.5$\pm$$\le$0.1 &  &  & -0.9$\pm$0.1 &  &  &  &  &  \\
J055931.30+201959.8 & -22.0$\pm$0.5 &  &  &  &  &  &  & 0.7$\pm$$\le$0.1 & -0.4$\pm$$\le$0.1* & -0.9$\pm$0.2 &  &  &  &  &  &  \\
J060028.33+225621.9 & -33.6$\pm$0.8 & -2.1$\pm$$\le$0.1 & -1.6$\pm$$\le$0.1 & -1.4$\pm$$\le$0.1 & 0.8$\pm$0.1 &  &  & 1.0$\pm$$\le$0.1 & -0.2$\pm$$\le$0.1 & 0.4$\pm$0.1* &  &  &  &  & -0.5$\pm$$\le$0.1 & -0.7$\pm$0.1 \\
J060050.65+194920.1 &  &  &  &  & 0.6$\pm$0.1 &  &  & 1.5$\pm$0.1 &  &  &  &  & -0.5$\pm$0.2 & -1.8$\pm$0.1 & -1.3$\pm$$\le$0.1 & -0.9$\pm$$\le$0.1 \\
J060056.64+022949.9 & -19.2$\pm$0.6 &  &  &  &  &  &  & 0.5$\pm$0.1 &  &  & -0.2$\pm$$\le$0.1 &  &  &  &  &  \\
J060404.09+214057.1 & -34.3$\pm$1.3 &  &  &  & 0.5$\pm$$\le$0.1 & 0.2$\pm$0.1 &  & 0.4$\pm$$\le$0.1 &  &  &  &  &  &  &  &  \\
J060414.76+240402.4 & -16.9$\pm$0.7 &  &  &  & 0.7$\pm$$\le$0.1 & 0.2$\pm$$\le$0.1 &  & 2.0$\pm$0.2 &  & 0.4$\pm$$\le$0.1 &  &  &  &  &  &  \\
J060442.49+300803.0 & -45.0$\pm$0.6 &  &  &  &  &  &  &  &  &  &  &  &  &  &  &  \\
J060528.79+121434.1 &  &  &  &  &  &  &  & 0.8$\pm$$\le$0.1 &  &  & -1.7$\pm$$\le$0.1 &  &  &  &  &  \\
J060559.62-055408.8 & -23.7$\pm$$\le$0.1 & 1.0$\pm$0.1 & 1.9$\pm$$\le$0.1 & 2.9$\pm$0.1 & 0.4$\pm$$\le$0.1 & 0.2$\pm$$\le$0.1 & 0.2$\pm$$\le$0.1* & 0.9$\pm$0.1 & -0.2$\pm$0.1 & 1.0$\pm$0.1 & -0.3$\pm$$\le$0.1 &  &  &  &  &  \\
J060708.29+073046.0 & -24.8$\pm$0.8 &  &  &  &  &  &  & 0.4$\pm$0.1 &  &  &  &  &  &  &  &  \\
J060723.53+210721.1 & -8.0$\pm$0.9 & 0.6$\pm$$\le$0.1 & 1.3$\pm$0.1 & 1.8$\pm$0.1 & 0.3$\pm$$\le$0.1 &  &  & 0.4$\pm$0.1 & -0.3$\pm$0.1 &  &  &  &  &  &  &  \\
J060741.70+292253.6 &  &  &  &  &  &  &  & 1.0$\pm$0.1* &  &  &  &  & -0.3$\pm$$\le$0.1* & -0.9$\pm$0.1* & -0.8$\pm$$\le$0.1 & -0.9$\pm$0.2 \\
J060926.79+245519.8 &  & -1.1$\pm$$\le$0.1 & -1.3$\pm$0.1 & -1.2$\pm$0.1 & 0.5$\pm$0.1 &  &  & 1.1$\pm$0.2 &  &  & -0.4$\pm$0.1* & -0.2$\pm$$\le$0.1* &  &  &  &  \\
J061141.94+203339.3 & -27.4$\pm$1.0 &  &  &  & 0.7$\pm$0.1 & 0.2$\pm$$\le$0.1 &  & 1.1$\pm$0.1 & 0.0$\pm$$\le$0.1 &  &  &  &  &  &  &  \\
J061159.65+282934.6 & -12.3$\pm$0.8 & -2.6$\pm$0.2 & -2.1$\pm$$\le$0.1 & -1.9$\pm$$\le$0.1 & 0.3$\pm$$\le$0.1 & 0.2$\pm$$\le$0.1 &  & 1.2$\pm$0.1 &  &  & -0.2$\pm$$\le$0.1* &  &  &  &  &  \\
J061235.99+181046.6 & -23.4$\pm$0.1 &  &  &  & 0.5$\pm$$\le$0.1 & 0.3$\pm$$\le$0.1 &  & 0.4$\pm$$\le$0.1 &  &  &  &  &  &  &  &  \\
J061312.69+152036.5 &  & -2.1$\pm$0.2 & -2.6$\pm$$\le$0.1 & -2.7$\pm$0.1 & 0.8$\pm$0.1 &  &  &  &  &  &  &  &  &  &  &  \\
J061337.26-062501.6 & -11.2$\pm$0.7 & -1.3$\pm$0.1* & -1.1$\pm$0.1* & -0.9$\pm$0.1* & 0.3$\pm$$\le$0.1 &  & 0.1$\pm$$\le$0.1* & 1.0$\pm$0.2 & -0.1$\pm$$\le$0.1 &  & -1.3$\pm$0.2 &  &  &  &  &  \\
J061400.02+274212.2 & -5.0$\pm$0.4 & 1.9$\pm$0.1 & 3.0$\pm$0.3 & 1.8$\pm$1.0 & 0.4$\pm$0.1* &  &  & 1.1$\pm$$\le$0.1 & -0.3$\pm$0.1 &  &  &  &  &  &  &  \\
J061804.84+231907.2 &  & -3.0$\pm$0.4 & -2.1$\pm$0.4 & -1.7$\pm$0.4 & 1.1$\pm$0.2 &  &  &  &  &  &  &  & -7.0$\pm$0.7 & -7.8$\pm$$\le$0.1 & -8.5$\pm$1.0 & -6.1$\pm$$\le$0.1 \\
J061839.06+171801.5 &  & -2.0$\pm$0.2 & -2.8$\pm$$\le$0.1 & -2.4$\pm$0.1 & 0.6$\pm$$\le$0.1 &  &  &  &  &  &  &  & -4.9$\pm$0.5 & -13.0$\pm$1.1 & -6.5$\pm$0.1 & -4.4$\pm$0.2 \\
J061946.60+210904.1 & -65.2$\pm$1.6 & -5.2$\pm$$\le$0.1 & -4.6$\pm$0.3 & -3.9$\pm$$\le$0.1 & 0.2$\pm$$\le$0.1 &  &  & 1.2$\pm$0.2 & -0.6$\pm$$\le$0.1* & 0.5$\pm$$\le$0.1 &  &  &  &  &  &  \\
J062029.71+040604.6 & -22.9$\pm$0.6 &  &  &  & 0.4$\pm$$\le$0.1 & 0.2$\pm$$\le$0.1 &  & 0.7$\pm$$\le$0.1 &  & 0.7$\pm$0.1 &  &  &  &  &  &  \\
J062105.17+194802.7 &  &  &  &  &  &  &  & 1.1$\pm$0.2 &  &  &  &  &  & -0.9$\pm$$\le$0.1 & -2.3$\pm$0.1 & -1.3$\pm$$\le$0.1 \\
J062245.04+213111.8 & -29.3$\pm$$\le$0.1 &  &  &  & 0.2$\pm$$\le$0.1 &  & 0.1$\pm$$\le$0.1* & 0.6$\pm$$\le$0.1 & -0.3$\pm$0.1 & 0.6$\pm$0.1* &  &  &  &  &  &  \\
J062306.42+224659.8 & -24.3$\pm$0.8 & 1.3$\pm$0.1 & 1.2$\pm$$\le$0.1 & 3.9$\pm$$\le$0.1 & 0.5$\pm$$\le$0.1 &  &  & 1.0$\pm$0.1 & -0.2$\pm$$\le$0.1 & 0.7$\pm$$\le$0.1 &  &  & -0.1$\pm$$\le$0.1* &  & -0.3$\pm$$\le$0.1* &  \\
J063151.56+045417.4 & -43.2$\pm$1.5 &  &  &  & 0.4$\pm$$\le$0.1 &  &  & 0.4$\pm$0.1 &  & 0.1$\pm$$\le$0.1* &  &  & -0.3$\pm$$\le$0.1* & -0.9$\pm$0.1 & -1.0$\pm$$\le$0.1 & -0.7$\pm$$\le$0.1 \\
J063622.64+184131.5 &  &  &  &  &  &  &  & 1.6$\pm$$\le$0.1 &  &  &  &  &  &  &  &  \\
J064105.87+092255.6 & -30.1$\pm$1.1 & 0.9$\pm$$\le$0.1* & 0.5$\pm$$\le$0.1* & 2.3$\pm$0.2* &  &  &  &  & -0.2$\pm$$\le$0.1 & 0.1$\pm$$\le$0.1* & -3.0$\pm$0.1 &  &  &  & -4.5$\pm$0.1 &  \\
J064656.42+011640.5 & -166.5$\pm$2.7 & -5.1$\pm$0.2 & -4.6$\pm$$\le$0.1 & -5.4$\pm$0.1 & 0.6$\pm$0.1 &  &  & 1.1$\pm$0.2 & -0.5$\pm$0.1* &  & -0.5$\pm$0.1* & -0.2$\pm$$\le$0.1* &  &  &  &  \\
J065334.09+010608.4 & -36.4$\pm$0.3 & -2.9$\pm$$\le$0.1 & -2.7$\pm$0.7 & -3.3$\pm$0.1 & 0.3$\pm$0.1 &  &  & 0.6$\pm$$\le$0.1 & -0.4$\pm$$\le$0.1 & 0.4$\pm$0.1 &  &  &  &  &  &  \\
J071043.86+060007.9 & -56.1$\pm$0.3 &  &  &  &  &  &  & 1.5$\pm$0.1 & -0.4$\pm$$\le$0.1* &  &  &  &  &  &  &  \\
J071652.68+352137.0 & -6.3$\pm$0.9 & 1.1$\pm$$\le$0.1 & 2.5$\pm$0.3 & 2.1$\pm$0.2 & 0.1$\pm$0.1 &  & 0.1$\pm$$\le$0.1 & 1.0$\pm$0.1 & -0.3$\pm$0.1 &  &  &  &  &  &  &  \\
J072715.16+081622.5 & -11.0$\pm$$\le$0.1 &  &  &  &  &  &  & 1.9$\pm$0.1 &  &  &  &  &  &  &  &  \\
J075320.02+154647.6 & -15.9$\pm$0.5 & 1.6$\pm$0.4 & 2.0$\pm$$\le$0.1 & 6.0$\pm$$\le$0.1 &  &  &  & 1.0$\pm$0.1 & -0.1$\pm$0.1 &  &  &  &  &  &  &  \\
J102441.62+354400.2 &  &  &  &  &  &  &  & 0.4$\pm$0.1 &  &  &  &  &  &  &  &  \\
J163906.42+094755.3 &  & 1.5$\pm$$\le$0.1 & 2.5$\pm$$\le$0.1 & 2.6$\pm$$\le$0.1 &  &  &  & 0.4$\pm$0.1 &  &  &  &  &  &  &  &  \\
J200556.10+333425.8 & -12.5$\pm$0.4 &  &  &  &  &  &  & 1.5$\pm$0.1 &  &  & -1.5$\pm$0.4 &  &  &  &  &  \\
J202654.98+393745.9 & -19.8$\pm$$\le$0.1 & 0.7$\pm$$\le$0.1 & 1.9$\pm$0.1 & 3.7$\pm$0.1 & 0.4$\pm$$\le$0.1 &  & 0.4$\pm$$\le$0.1* & 1.1$\pm$0.1 & -0.4$\pm$0.1 &  & -0.4$\pm$$\le$0.1 &  & -0.7$\pm$0.1* & -2.0$\pm$0.1* &  &  \\
J203546.48+354719.6 & -22.3$\pm$0.1 & -2.8$\pm$$\le$0.1 & -2.6$\pm$0.1 & -1.7$\pm$$\le$0.1 & 0.5$\pm$$\le$0.1 & 0.1$\pm$$\le$0.1 & 0.5$\pm$0.6 & 1.1$\pm$0.1 & -1.7$\pm$$\le$0.1 & 0.5$\pm$0.1 &  &  &  &  &  &  \\
J204106.79+423111.6 & -18.8$\pm$0.4 &  &  &  & 0.4$\pm$$\le$0.1 & 0.1$\pm$$\le$0.1 &  & 1.6$\pm$$\le$0.1 & -0.1$\pm$$\le$0.1 & -0.2$\pm$0.1* &  &  &  &  &  &  \\
J205229.00+383517.7 & -17.5$\pm$0.5 & 1.1$\pm$0.1 & 1.3$\pm$$\le$0.1 & 3.0$\pm$$\le$0.1 &  &  &  & 0.9$\pm$$\le$0.1 & -0.4$\pm$$\le$0.1 &  &  &  &  &  &  &  \\
J213231.99+385754.9 & -14.3$\pm$0.4 & 1.0$\pm$$\le$0.1 & 1.7$\pm$$\le$0.1 & 3.4$\pm$$\le$0.1 & 0.2$\pm$$\le$0.1 &  & 0.1$\pm$$\le$0.1 & 1.2$\pm$0.1 & -0.1$\pm$$\le$0.1 &  &  &  &  &  &  &  \\
J224515.96+563739.7 & -4.4$\pm$0.2 &  &  &  & 0.5$\pm$0.1 &  &  &  &  & 0.3$\pm$0.1 &  &  &  &  &  &  

\end{longtable}
\end{landscape}  

%

\onecolumn
\begin{landscape}
\begin{longtable}{lllllll}
\caption{List of HAeBe stars identified from this study. Spectral type obtained from template matching with MILES library spectra, extinction values estimated using the EW of DIB spectral feature, distance estimates of these stars obtained from \citet{2021yCat.1352....0B}, estimated age, mass, and mass accretion rate for the individual stars from this study are provided. For stars in which the A$_V$ is not obtained using the DIB spectral feature, we have retrieved extinction values from \citet{2019ApJ...887...93G}. The $A_{V}$ values obtained from Green's map are regarded as lower limits to the extinction and are indicated with an asterisk ($*$).}
\label{tab:A2}\\
\hline
LAMOST\_ID & Spectral type & Distance & $A_{V}$ & Age & Mass & log($\dot{M}_{acc}$) \\ 
        ~ & ~ & (mag) & (pc) & (Myr) & ($M_{\odot}$) & ($M_{\odot}yr^{-1}$) \\ \hline

\endfirsthead
\multicolumn{7}{c}%
{\bfseries Table \thetable\ continued from previous page} \\
\hline
LAMOST\_ID & Spectral type & Distance & A$_{V}$ & Age & Mass & log($\dot{M}_{acc}$) \\ 
       ~ & ~ & (pc) & (mag) & (Myr) & ($M_{\odot}$) & ($M_{\odot}yr^{-1}$) \\ \hline
\endhead
\hline
\endfoot
\endlastfoot
J024245.73+563110.5 &  & 2250$\pm$158 & 1.4$\pm$0.2 & 14.8$\pm$1.50 & 1.46$\pm$0.07 &  \\
J025146.96+554201.3 & A2V & 2638$\pm$118 & 2.7$\pm$0.7 & 0.77$\pm$0.08 & 4.42$\pm$0.15 & -6.18$\pm$0.25 \\
J032832.62+511354.4 & A0III & 1928$\pm$51 & 2.7$\pm$0.2 & 0.78$\pm$0.04 & 4.31$\pm$0.09 & -6.56$\pm$0.11 \\
J032907.54+570133.7 &  & 2270$\pm$133 & 2.5$\pm$0.1 & 0.23$\pm$0.02 & 4.9$\pm$0.12 &  \\
J033900.56+294145.7 & A0V & 389$\pm$4 & 1.0$\pm$0.4 & 8.63$\pm$0.48 & 1.78$\pm$0.01 & -6.98$\pm$0.21 \\
J034109.13+314437.8 &  & 298$\pm$4 & 1.0$\pm$0.1 & 0.78$\pm$0.05 & 0.29$\pm$0.01 &  \\
J034255.96+315841.7 & A3V & 305$\pm$2 & 1.1$\pm$0.2 & 0.67$\pm$0.03 & 0.49$\pm$0.01 & -8.07$\pm$0.04 \\
J035553.38+363840.4 & F2 & 3303$\pm$733 & 0.6$\pm$0.1* &  &  &  \\
J035952.31+481339.8 & B2IV & 3067$\pm$127 & 2.8$\pm$0.4 &  &  &  \\
J040058.11+572247.7 & B5 & 2699$\pm$102 & 1.6$\pm$0.3 & 1.35$\pm$0.10 & 3.54$\pm$0.1 & -6.09$\pm$0.11 \\
J040340.25+344537.6 &  & 4683$\pm$1239 & 0.6$\pm$0.1* &  &  &  \\
J040559.62+295638.1 & A0III & 338$\pm$2 & 0.9$\pm$0.2 & 4.4$\pm$0.18 & 1.75$\pm$0.01 & -8.59$\pm$0.1 \\
J041150.68+503619.6 & B1III & 2648$\pm$194 & 2.6$\pm$0.2 & 0.1$\pm$$\le$0.01 & 8.91$\pm$0.57 & -5.63$\pm$0.1 \\
J041805.77+533707.2 & B6IV & 3847$\pm$233 & 3.3$\pm$0.5 & 0.1$\pm$$\le$0.01 & 9.41$\pm$0.36 & -5.18$\pm$0.24 \\
J042622.24+440031.4 & B2III & 2391$\pm$97 & 2.6$\pm$0.5 &  &  &  \\
J042853.11+483132.4 & A2 & 1719$\pm$77 & 2.6$\pm$0.2 & 5.29$\pm$0.23 & 2.14$\pm$0.03 & -6.9$\pm$0.24 \\
J043000.64+422259.6 & B2IV & 2465$\pm$81 & 1.9$\pm$0.2 & 0.76$\pm$0.05 & 4.41$\pm$0.1 & -5.82$\pm$0.13 \\
J043718.84+450641.9 & F5V & 1104$\pm$24 & 2.0$\pm$0.1 & 8.92$\pm$0.17 & 1.78$\pm$0.03 & -7.34$\pm$0.09 \\
J045918.31+284057.7 &  & 4768$\pm$1793 & 1.6$\pm$0.2* & 2.59$\pm$2.42 & 2.6$\pm$0.63 &  \\
J050303.79+422144.6 & A0 & 6137$\pm$1088 & 2.8$\pm$0.4 & 1.38$\pm$0.49 & 3.5$\pm$0.45 & -5.67$\pm$0.25 \\
J050455.66+143155.5 &  & 4209$\pm$1816 & 0.8$\pm$0.1* &  &  &  \\
J050924.57+520007.1 & A1V & 1763$\pm$49 & 2.3$\pm$0.3 & 2.73$\pm$0.15 & 2.74$\pm$0.07 & -6.63$\pm$0.16 \\
J051039.20+305419.6 & A3 & 2067$\pm$177 & 2.1$\pm$0.3 &  &  &  \\
J051133.52+334849.7 &  & 3940$\pm$1265 & 2.2$\pm$0.2 &  &  &  \\
J051159.17+342843.0 &  & 3679$\pm$1297 & 1.7$\pm$0.3* & 13.4$\pm$4.85 & 1.5$\pm$0.37 &  \\
J051425.20+411310.7 & B8III & 1218$\pm$74 & 0.5$\pm$0.1* &  &  &  \\
J051439.95+324030.5 & A2V & 1279$\pm$29 & 2.9$\pm$0.6 &  &  &  \\
J051634.39+402856.1 &  & 4419$\pm$931 & 2.5$\pm$0.4 & 3.4$\pm$1.38 & 2.5$\pm$0.19 &  \\
J051727.81+344343.0 &  & 2203$\pm$71 & 1.2$\pm$0.1 &  &  &  \\
J052017.50+391657.8 & B8III & 2263$\pm$156 & 2.4$\pm$0.3 & 0.47$\pm$0.07 & 5.31$\pm$0.24 & -5.94$\pm$0.16 \\
J052019.20+320817.0 & B2IV & 2961$\pm$113 & 2.4$\pm$0.4 &  &  &  \\
J052033.42+330857.6 & A7m & 2984$\pm$1141 & 1.2$\pm$0.2* &  &  &  \\
J052050.38+315637.5 & B8 & 3722$\pm$193 & 1.0$\pm$0.1 &  &  &  \\
J052058.21+041827.6 & B9V & 5845$\pm$1588 & 0.2$\pm$$\le$0.1* &  &  &  \\
J052135.30-045329.2 & A2V & 370$\pm$4 & 1.6$\pm$0.1 &  &  &  \\
J052250.18+413711.9 & A1V & 6166$\pm$2755 & 1.6$\pm$0.2* & 8.5$\pm$3.82 & 1.74$\pm$0.57 & -7.42$\pm$0.35 \\
J052322.02+281011.3 &  & 3843$\pm$1690 & 1.5$\pm$0.7 &  &  &  \\
J052330.87+051726.8 & A7V & 3391$\pm$1077 & 1.8$\pm$0.3 &  &  &  \\
J052419.17+393734.4 &  & 6050$\pm$2011 & 1.6$\pm$0.2* & 0.1$\pm$$\le$0.01 & 0.93$\pm$0.21 &  \\
J052647.47+114752.5 &  & 3558$\pm$1223 & 1.4$\pm$0.2* &  &  &  \\
J052707.42+341822.7 & A3V & 2378$\pm$226 & 1.6$\pm$0.2* &  &  &  \\
J052747.93+084225.8 &  & 3161$\pm$746 & 0.8$\pm$0.1* &  &  &  \\
J052919.14+341747.1 & B8 & 2588$\pm$145 & 2.7$\pm$0.3 &  &  &  \\
J052926.22+351610.5 & A0V & 1933$\pm$63 & 2.3$\pm$0.1 &  &  &  \\
J052947.46+332915.5 & A0V & 1846$\pm$105 & 2.5$\pm$0.5 &  &  &  \\
J052948.05-002343.4 & A2V & 402$\pm$5 & 1.0$\pm$0.1 &  &  &  \\
J053024.01+373010.2 &  & 2934$\pm$210 & 2.5$\pm$0.2 & 1.59$\pm$0.23 & 3.4$\pm$0.19 &  \\
J053123.43+383157.1 & B5V & 4219$\pm$485 & 1.8$\pm$0.4 & 0.29$\pm$0.04 & 7.0$\pm$0.36 & -6.82$\pm$0.3 \\
J053148.50+152211.7 & A0III & 4473$\pm$1175 & 0.8$\pm$0.1 &  &  &  \\
J053200.30-045553.8 &  & 347$\pm$6 & 0.5$\pm$0.2 &  &  &  \\
J053209.94-024946.7 & F5V & 347$\pm$2 & 1.0$\pm$0.1 & 6.63$\pm$1.67 & 1.95$\pm$0.18 & -7.81$\pm$0.01 \\
J053227.09+090006.1 &  & 5493$\pm$1769 & 0.6$\pm$0.1* & 14.0$\pm$4.95 & 1.49$\pm$0.33 &  \\
J053625.75-021000.8 &  & 6806$\pm$1760 & 0.4$\pm$0.1* &  &  &  \\
J053637.36+330003.9 & A9V & 1914$\pm$104 & 1.0$\pm$0.2 &  &  &  \\
J053857.22+373003.2 & A0V & 1583$\pm$61 & 2.3$\pm$0.4 &  &  &  \\
J053918.09+361716.2 & A2VI & 1530$\pm$80 & 2.7$\pm$0.1 & 3.87$\pm$0.43 & 2.44$\pm$0.21 & -7.51$\pm$0.08 \\
J054109.42+360836.8 & A2V & 1780$\pm$179 & 1.0$\pm$0.1 &  &  &  \\
J054113.78+273938.5 &  & 2003$\pm$637 & 0.9$\pm$0.1 &  &  &  \\
J054252.77+114401.9 &  & 3193$\pm$966 & 0.8$\pm$0.1* &  &  &  \\
J054329.26+000458.8 & A6IV-V & 429$\pm$5 & 1.2$\pm$0.2 & 9.2$\pm$1.28 & 1.71$\pm$0.13 & -7.15$\pm$0.05 \\
J054424.58+342802.7 & B9IV & 5814$\pm$2008 & 2.6$\pm$0.3 &  &  &  \\
J054431.95+320449.2 & F0 & 1371$\pm$50 & 2.4$\pm$0.1 &  &  &  \\
J054454.86+215736.4 & F4V & 1287$\pm$38 & 2.3$\pm$0.2 & 7.52$\pm$0.37 & 1.9$\pm$0.04 & -7.1$\pm$0.14 \\
J054649.95+220630.2 &  & 3662$\pm$1154 & 2.7$\pm$0.1 &  &  &  \\
J054658.35+304125.4 & A2 & 1726$\pm$104 & 2.4$\pm$0.4 &  &  &  \\
J055054.77+201447.6 & B8 & 1381$\pm$28 & 2.8$\pm$0.4 & 1.73$\pm$0.12 & 3.3$\pm$0.14 &  \\
J055419.97-003838.1 &  & 4821$\pm$1468 & 1.3$\pm$0.2* & 13.2$\pm$3.23 & 1.61$\pm$0.86 &  \\
J055455.22-024334.0 &  & 4678$\pm$1080 & 1.3$\pm$0.2* &  &  &  \\
J055530.50+115420.6 & A2V & 5110$\pm$1167 & 2.7$\pm$0.1 &  &  &  \\
J055547.52+174350.0 & A3m & 1667$\pm$58 & 2.3$\pm$0.2 & 4.47$\pm$0.29 & 2.32$\pm$0.06 & -6.59$\pm$0.12 \\
J055831.50+264236.0 & B8V & 4836$\pm$665 & 2.3$\pm$0.4 & 0.77$\pm$0.20 & 4.41$\pm$0.42 & -5.96$\pm$0.21 \\
J055838.98+201108.5 & B6IV & 1796$\pm$73 & 2.3$\pm$0.5 & 3.27$\pm$0.26 & 2.65$\pm$0.08 &  \\
J055910.90+120215.0 &  & 5598$\pm$1559 & 1.3$\pm$0.1 &  &  &  \\
J055931.30+201959.8 & A3 & 1626$\pm$64 & 1.7$\pm$0.1 &  &  &  \\
J060028.33+225621.9 & B2III & 4398$\pm$528 & 2.2$\pm$0.5 & 0.32$\pm$0.09 & 5.78$\pm$0.5 & -6.4$\pm$0.21 \\
J060050.65+194920.1 &  & 4454$\pm$1505 & 0.7$\pm$0.0 & 7.6$\pm$1.40 & 1.8$\pm$1.04 &  \\
J060056.64+022949.9 & A0III & 5343$\pm$1577 & 1.5$\pm$0.2* &  &  &  \\
J060404.09+214057.1 & A0V & 2346$\pm$334 & 2.5$\pm$0.4 &  &  &  \\
J060414.76+240402.4 & B5V & 4736$\pm$337 & 3.8$\pm$0.1 &  &  &  \\
J060442.49+300803.0 & A2V & 3283$\pm$632 & 2.3$\pm$0.3* &  &  &  \\
J060528.79+121434.1 &  & 4371$\pm$1130 & 2.2$\pm$0.1 &  &  &  \\
J060559.62-055408.8 & A1V & 796$\pm$11 & 2.7$\pm$0.2 &  &  &  \\
J060708.29+073046.0 & A2 & 3471$\pm$777 & 0.5$\pm$0.1* &  &  &  \\
J060723.53+210721.1 & F5V & 2332$\pm$489 & 1.9$\pm$0.1 & 6.59$\pm$3.15 & 1.97$\pm$0.36 & -7.42$\pm$0.19 \\
J060741.70+292253.6 &  & 4890$\pm$1222 & 2.4$\pm$0.4* & 5.1$\pm$2.38 & 2.24$\pm$0.42 &  \\
J060926.79+245519.8 & B8 & 1672$\pm$94 & 2.3$\pm$0.8 & 3.48$\pm$0.35 & 2.45$\pm$0.1 &  \\
J061047.12-061250.7 & B9 & 846$\pm$15 & 4.7$\pm$0.7* & 3.87$\pm$0.10 & 2.4$\pm$0.03 &  \\
J061141.94+203339.3 & A0V & 2336$\pm$275 & 2.4$\pm$0.5 & 1.01$\pm$0.09 & 4.63$\pm$0.1 & -6.45$\pm$0.55 \\
J061159.65+282934.6 & B8V & 5337$\pm$1180 & 2.6$\pm$0.4 & 1.44$\pm$0.71 & 3.57$\pm$0.59 & -6.46$\pm$0.25 \\
J061235.99+181046.6 & A0V: & 1944$\pm$131 & 2.3$\pm$0.3 &  &  &  \\
J061312.69+152036.5 &  & 2980$\pm$296 & 1.4$\pm$0.4 & 0.65$\pm$0.17 & 5.0$\pm$0.38 &  \\
J061337.26-062501.6 & A2 & 834$\pm$11 & 1.9$\pm$0.1 & 3.42$\pm$0.07 & 2.53$\pm$0.02 & -6.69$\pm$0.09 \\
J061400.02+274212.2 & F4V & 1703$\pm$192 & 1.7$\pm$0.1 & 0.1$\pm$$\le$0.01 & 3.6$\pm$1.19 & -6.12$\pm$0.04 \\
J061804.84+231907.2 &  & 3051$\pm$482 & 2.5$\pm$0.8 & 0.77$\pm$0.26 & 4.33$\pm$0.51 &  \\
J061839.06+171801.5 &  & 2469$\pm$251 & 2.4$\pm$0.4 & 3.8$\pm$0.69 & 2.43$\pm$0.17 &  \\
J061946.60+210904.1 & B5V & 1715$\pm$45 & 1.9$\pm$0.1 & 0.26$\pm$0.02 & 6.47$\pm$0.12 & -5.18$\pm$0.08 \\
J062029.71+040604.6 & B5V & 6055$\pm$672 & 2.4$\pm$0.1 &  &  &  \\
J062105.17+194802.7 &  & 3238$\pm$876 & 2.9$\pm$0.5 &  &  &  \\
J062245.04+213111.8 & A1V & 1756$\pm$62 & 2.2$\pm$0.1 &  &  &  \\
J062306.42+224659.8 & A0V & 1577$\pm$82 & 2.4$\pm$0.1 & 8.6$\pm$0.55 & 1.71$\pm$0.03 & -7.14$\pm$0.18 \\
J063151.56+045417.4 & A0V & 1412$\pm$35 & 2.5$\pm$0.1 &  &  &  \\
J063622.64+184131.5 &  & 6072$\pm$2007 & 1.2$\pm$0.1 &  &  &  \\
J063856.02+542940.4 &  & 3433$\pm$802 & 0.2$\pm$$\le$0.1* & 6.09$\pm$3.21 & 1.9$\pm$0.32 &  \\
J064105.87+092255.6 & A2 & 709$\pm$10 & 1.0$\pm$0.0 &  &  &  \\
J064656.42+011640.5 & B5V & 1558$\pm$129 & 0.8$\pm$0.0 & 4.1$\pm$0.63 & 2.42$\pm$0.15 & -7.28$\pm$0.09 \\
J065334.09+010608.4 & B5V & 3756$\pm$185 & 2.4$\pm$0.5 &  &  &  \\
J071043.86+060007.9 & B9 & 542$\pm$9 & 0.1$\pm$$\le$0.1* &  &  &  \\
J071652.68+352137.0 & F5V & 2259$\pm$103 & 0.6$\pm$0.2 & 9.05$\pm$0.17 & 1.73$\pm$0.05 & -7.37$\pm$0.19 \\
J072715.16+081622.5 & B9III & 1386$\pm$849 & 0.1$\pm$$\le$0.1* &  &  &  \\
J075320.02+154647.6 & A2 & 3065$\pm$210 & 1.5$\pm$0.3 & 1.55$\pm$0.19 & 3.43$\pm$0.14 & -6.01$\pm$0.18 \\
J102441.62+354400.2 &  & 5817$\pm$2145 & 0.9$\pm$0.2 & 16.6$\pm$5.01 & 1.37$\pm$0.8 &  \\
J163906.42+094755.3 &  & 1220$\pm$26 & 0.9$\pm$0.3 & 8.93$\pm$0.31 & 1.76$\pm$0.06 &  \\
J200556.10+333425.8 & A2 & 1642$\pm$34 & 1.0$\pm$0.2 & 3.0$\pm$0.15 & 2.8$\pm$0.07 & -6.63$\pm$0.07 \\
J202654.98+393745.9 & F0 & 1721$\pm$68 & 2.3$\pm$0.2 & 1.62$\pm$0.11 & 3.13$\pm$0.07 & -7.82$\pm$0.1 \\
J203546.48+354719.6 & B2III & 2321$\pm$80 & 3.0$\pm$0.1 & 0.31$\pm$0.03 & 5.96$\pm$0.15 & -5.32$\pm$0.08 \\
J204106.79+423111.6 & A0V & 954$\pm$49 & 2.4$\pm$0.2 &  &  &  \\
J205229.00+383517.7 & A2V & 869$\pm$10 & 1.9$\pm$0.1 &  &  &  \\
J213231.99+385754.9 & A2V & 899$\pm$13 & 1.3$\pm$0.2 & 8.66$\pm$0.50 & 1.68$\pm$0.02 & -7.02$\pm$0.26 \\
J224515.96+563739.7 & B2III & 3574$\pm$138 & 2.7$\pm$0.4 & 0.1$\pm$$\le$0.01 & 10.7$\pm$0.65 & -5.95$\pm$0.08
\end{longtable}
\end{landscape}


\bsp	
\label{lastpage}
\end{document}